\documentclass[american,aps,prb,reprint,superscriptaddress,nofootinbib]{revtex4-2}
\usepackage[T1]{fontenc}
\usepackage{color}
\usepackage{babel}
\usepackage{amsmath}
\usepackage{amssymb}
\usepackage{graphicx}
\usepackage{esint}
\usepackage[unicode=true,pdfusetitle,
 bookmarks=true,bookmarksnumbered=false,bookmarksopen=false,
 breaklinks=false,pdfborder={0 0 0},pdfborderstyle={},backref=false,colorlinks=true]
 {hyperref}
\hypersetup{
 citecolor=blue,linkcolor=magenta,urlcolor=blue}

\makeatletter
\usepackage{xcolor}

\makeatother

\begin{document}
\global\long\def\i{\mathrm{i}}%
\global\long\def\e{\mathrm{e}}%
\global\long\def\d{\mathrm{d}}%
\global\long\def\bra#1{\left\langle #1\right|}%
\global\long\def\ket#1{\left|#1\right\rangle }%
\global\long\def\braket#1#2{\left\langle #1|#2\right\rangle }%
\global\long\def\ketbra#1#2{\left|#1\right\rangle \!\left\langle #2\right|}%
\global\long\def\Tr{\mathrm{Tr}}%

\title{Two-dimensional Thouless pumping in time-space crystalline structures}
\author{Y. Braver}
\thanks{These authors contributed equally}
\affiliation{Institute of Theoretical Physics and Astronomy, Vilnius University,
Saul\.{e}tekio 3, LT-10257 Vilnius, Lithuania}
\author{C.-h. Fan}
\thanks{These authors contributed equally}
\affiliation{Instytut Fizyki Teoretycznej, Uniwersytet Jagiello\'{n}ski, ulica
Profesora Stanis\l awa \L ojasiewicza 11, PL-30-348 Krak\'{o}w, Poland}
\affiliation{School of Physics and Optoelectronics, Xiangtan University, Hunan
411105, China}
\author{G. \v{Z}labys}
\thanks{These authors contributed equally}
\affiliation{Institute of Theoretical Physics and Astronomy, Vilnius University,
Saul\.{e}tekio 3, LT-10257 Vilnius, Lithuania}
\affiliation{Quantum Systems Unit, Okinawa Institute of Science and Technology
Graduate University, Onna, Okinawa 904-0495, Japan}
\author{E. Anisimovas}
\affiliation{Institute of Theoretical Physics and Astronomy, Vilnius University,
Saul\.{e}tekio 3, LT-10257 Vilnius, Lithuania}
\author{K. Sacha}
\affiliation{Instytut Fizyki Teoretycznej, Uniwersytet Jagiello\'{n}ski, ulica
Profesora Stanis\l awa \L ojasiewicza 11, PL-30-348 Krak\'{o}w, Poland}
\date{\today}
\begin{abstract}
Dynamics of a particle in a resonantly driven quantum well can be
interpreted as that of a particle in a crystal-like structure, with
the time playing the role of the coordinate. By introducing an adiabatically
varied phase in the driving protocol, we demonstrate a realization
of the Thouless pumping in such a time crystalline structure. Next,
we extend the analysis beyond a single quantum well by considering
a driven one-dimensional optical lattice, thereby engineering a 2D
time-space crystalline structure. Such a setup allows us to explore
adiabatic pumping in the spatial and the temporal dimensions separately,
as well as to simulate simultaneous time-space pumping.
\end{abstract}
\maketitle

\section{Introduction}

Recent developments in time-crystal research \citep{Sacha2017rev,Guo2020,SachaTC2020,GuoBook,Hannaford2022}
include the study of time crystalline structures --- closed quantum
systems that are driven by time-periodic external signals --- with
the aim to take advantage of thereby induced regular periodic repetition
observed in the time domain. In this way, time is endowed with properties
of an extra coordinate axis. We stress that in such scenarios \citep{Guo2013,Sacha15a},
the time-periodic structure is indeed imposed externally; here one
does not need to rely on the favorable role of particle interactions
for its spontaneous formation. Nevertheless, the manifestation of
the periodic regularity in the time domain is no less intriguing due
to its ability to simulate the familiar spatially periodic solid state
systems and phenomena of the condensed-matter realm. Recent systematic
studies resulted in a growing list of condensed-matter phases reproduced
or generalized in the time domain: Anderson and many-body localization,
Mott insulator, topological and other phases have been reported \citep{Guo2013,Sacha15a,Guo2016a,Guo2016,Mierzejewski2017,delande17,Giergiel2018,Lustig2018,Giergiel2018b,Peng2018,Peng2018a}.
Another extension was provided by the fresh proposal of \emph{time-space
crystalline} structures \citep{Zlabys2021}, see also \citep{Li2012,Gomez-Leon2013,Messer2018,Fujiwara2019,Cao2020,Gao2021,Martinez2021,Chakraborty2022},
that combine periodicity in time and in space. Viewed as an introduction
of synthetic dimensions, such time-space lattices pave the way to
potential doubling of the number of dimensions.

In this work, we explore a phenomenon that can be carried over from
conventional space crystals to the time or time-space crystalline
structures --- the Thouless pumping \citep{Thouless1983,Lohse2016pump,Nakajima2016,PetridesEtAl2018,LohseEtAl2018}.
As shown by Thouless, a suitably performed adiabatic variation of
the lattice parameters can lead to quantized particle transport along
the lattice. We demonstrate that an adiabatic variation of the external
driving can analogously lead to quantized particle motion in the temporal
dimension. To this end, we consider adiabatic pumping in a two-dimensional
(2D) time-space crystal realized by a resonantly driven optical lattice.
We study three possible processes: pumping in the temporal dimension,
spatial pumping, and simultaneous pumping in both dimensions. Remarkably,
a 2D Thouless pump may be used to study 4D quantum Hall effect \citep{LohseEtAl2018,Zilberberg2018,PetridesEtAl2018},
and hence the setup proposed here enables one to probe 4D physics
with just a driven system of a single spatial dimension.

\section{Model}

We base our demonstration of the time-space Thouless pumping on the
one-dimensional scaled Hamiltonian of the form
\begin{equation}
\hat{H}=\hat{h}(\hat{p}_{x},x)+\xi_{{\rm S}}(x,t)+\xi_{{\rm L}}(x,t|\varphi_{t}).\label{eq:H}
\end{equation}
The first term is the unperturbed spatial Hamiltonian,

\begin{equation}
\hat{h}(\hat{p}_{x},x)=\hat{p}_{x}^{2}-V_{{\rm S}}\cos^{2}(2x)-V_{{\rm L}}\cos^{2}(x+\varphi_{x}),\label{eq:h}
\end{equation}
which is typical for setups demonstrating the topological Thouless
pumping in the real space \citep{Lohse2016pump,Nakajima2016}. Here,
$\hat{p}_{x}^{2}$ is the momentum operator, $V_{{\rm S}}$ and $V_{{\rm L}}$
control, respectively, the depth of the ``short'' and the ``long''
optical lattices, while the relative phase $\varphi_{x}$ has to slowly
scan over a period of length $\pi$ to realize a pumping cycle. Throughout
this work, we use the recoil units for the energy $\hbar^{2}k_{{\rm L}}^{2}/2m$
and length $1/k_{{\rm L}}$, with $k_{{\rm L}}$ being the wave number
of laser beams that create the optical lattice and $m$ the particle
mass.

To be able to engineer the topological Thouless pumping in time, we
introduce the time-dependent perturbations

\vspace{-5mm}
\begin{subequations}
\label{eq:perturb}
\begin{align}
\xi_{{\rm S}}(x,t) & =\lambda_{{\rm S}}\sin^{2}(2x)\cos(2\omega t),\\
\xi_{{\rm L}}(x,t|\varphi_{t}) & =\lambda_{{\rm L}}\cos^{2}(2x)\cos(\omega t+\varphi_{t}).
\end{align}
\end{subequations}
The factors $\lambda_{{\rm S}}$ and $\lambda_{{\rm L}}$ denote the
overall strength of these perturbations. As we will demonstrate shortly,
the time-periodic dependencies $\cos(2\omega t)$ and $\cos(\omega t+\varphi_{t})$
enable us to introduce a pumping setup based on a periodic structure
in the time domain. The role of the phase shift $\varphi_{t}$ is
to allow for slowly changing the relative displacement between the
two emerging time lattices. In this work we choose the driving frequency
$\omega=s\Omega$, where the resonance number $s=2$, while $\Omega$
is the gap between neighboring energy bands of $\hat{h}$ which we
wish to couple by the external perturbation. The combination of perturbations
oscillating as $2\Omega$ and $4\Omega$ allows us to create a ring
of four sites (two cells with two sites per cell) in the temporal
direction. To ensure sufficient hopping strength between the sites
of the spatial lattice, we study highly excited states of $\hat{h}$
that occupy bands near the top of the spatial potential wells. The
corresponding value of $\Omega$ is easily determined by diagonalizing
$\hat{h}$. Since the phase $\varphi_{x}$ changes the spatial potential
and thus the unperturbed energy spectrum, we additionally fine-tune
the value of $\Omega$ (or $\omega$ directly) to keep the spatial
hopping adequate for all $\varphi_{x}$.

We note that the temporal phase $\varphi_{t}$ is not a parameter
of the unperturbed system, but rather a parameter of the perturbation.
Moreover, $\varphi_{x}$ and $\varphi_{t}$ change in time in the
Thouless pumping protocol. In order not to destroy the time crystalline
structure which is created by resonant time-periodic driving of the
system, $\varphi_{x}$ and $\varphi_{t}$ may not change appreciably
during a single period $T$ of the driving, i.e. $T(\d\varphi_{x,t}/\d t)\ll2\pi$.
Actually, when considering the Thouless pumping, a stronger condition
is assumed because not only the time crystal structure may not be
destroyed by the changes of $\varphi_{x}$ and $\varphi_{t}$, but
also the evolution of the system that forms this time crystalline
structure has to be adiabatic.

\begin{figure*}
\begin{centering}
\includegraphics{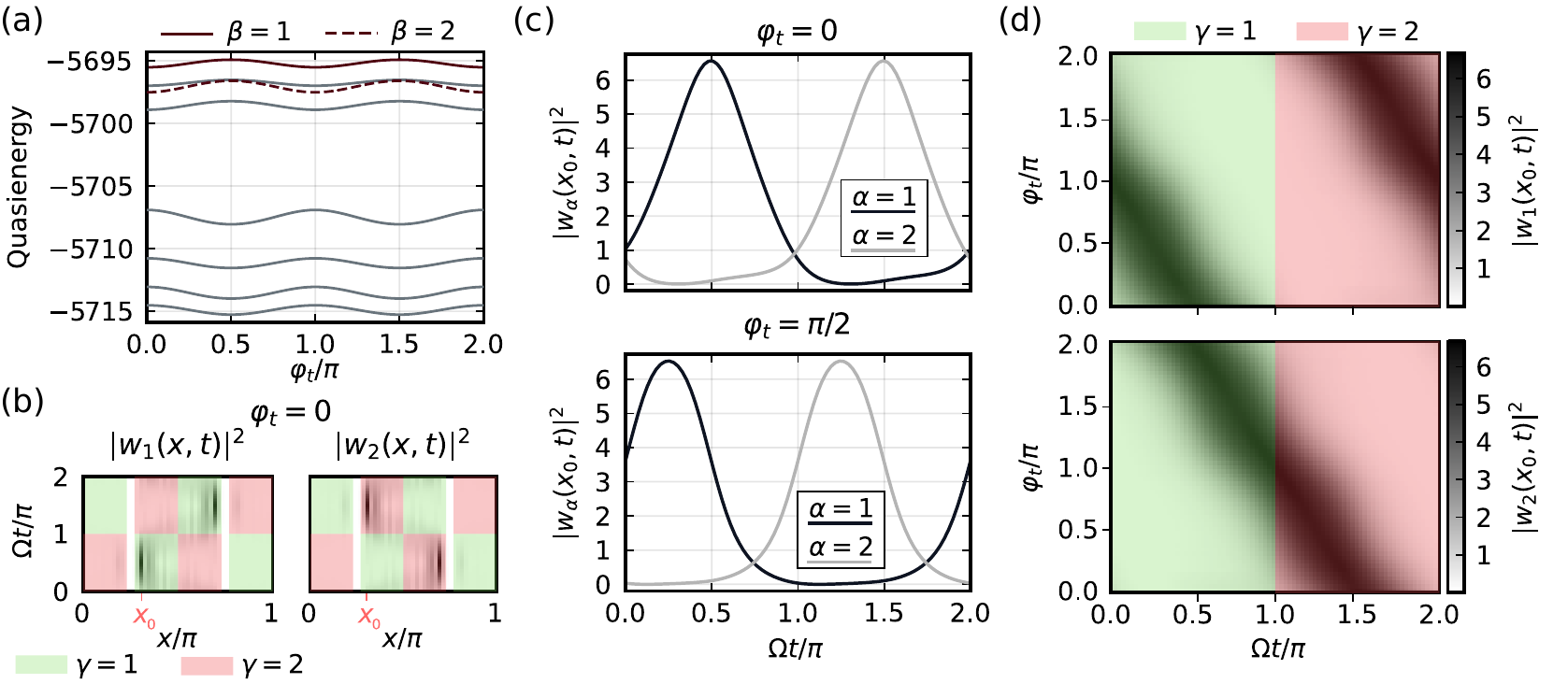}
\par\end{centering}
\caption{\label{fig:2D-pumping-time}Temporal adiabatic pumping in a 2D time-space
crystal with $s=2$ temporal cells and only one ($N=1$) spatial cell
that consists of two spatial sites. The following values of parameters
were used: $V_{{\rm S}}=7640$, $V_{{\rm L}}=2$, $\omega=410$, $s=2$,
$\lambda_{{\rm S}}=100$, $\lambda_{{\rm L}}=40$ and $\varphi_{x}=0$.
(a) Quasienergy levels $\varepsilon_{\beta}$ (see text for level
numbering convention) of the Floquet Hamiltonian $\hat{{\cal H}}$
versus the adiabatic phase $\varphi_{t}$. (b) The Wannier functions
$|w_{\alpha}(x,t)|^{2}$ at $\varphi_{t}=0$, represented by black
regions. The shaded areas (green and pink) indicate the extent of
the temporal cells ($\gamma=1,2$). (c) Wannier functions $|w_{\alpha}(x_{0},t)|^{2}$
(where $x_{0}=0.3\pi$) at $\varphi_{t}=0$ and $\varphi_{t}=\pi/2$.
(d) Change of the Wannier functions $|w_{\alpha}(x_{0},t)|^{2}$ as
$\varphi_{t}$ scans across a complete cycle of length $2\pi$. The
green and pink shaded areas indicate the extent of the two temporal
lattice cells ($\gamma=1,2$).}
\end{figure*}

Since the perturbation $\xi_{{\rm S}}+\xi_{{\rm L}}$ is time-periodic,
with the period $T=2\pi/\omega$ {[}see Eqs.~(\ref{eq:perturb}){]},
we approach the problem by introducing the Floquet Hamiltonian $\hat{{\cal H}}=\hat{H}-\i\partial_{t}$
and solving the eigenvalue problem $\hat{{\cal H}}u_{n}(x,t)=\varepsilon_{n}u_{n}(x,t)$
\citep{Shirley1965,Buchleitner2002}. Here, $\varepsilon_{n}$ is
the quasienergy of the $n$th eigenstate, while $u_{n}(x,t)$ is the
corresponding Floquet mode that respects temporal periodicity of the
perturbation, i.e. $u_{n}(x,t)=u_{n}(x,t+2\pi/\omega)$. A general
solution of the Schr\"{o}dinger equation can be represented as a superposition
of states $\Psi_{n}(x,t)=\e^{-\i\varepsilon_{n}t}u_{n}(x,t).$ In
our simulations we consider a finite number of spatial cells ($N=1$
or $N=2$) which leads to the Hamiltonian being defined on $x\in[0,N\pi)$,
and we always assume periodic boundary conditions. All the details
of the diagonalization procedure are covered in Appendix A.

\section{Simulations}

In the following sections, we present the results of the simulations,
starting with the Thouless pumping in time. Next, we consider pumping
in space, which, however, is performed in a time-space crystal rather
than a conventional space crystal. Finally, we study simultaneous
pumping in both the temporal and the spatial dimensions.

A very helpful way to illustrate Thouless pumping is to show how Wannier
states are transported with a change of an adiabatic parameter \citep{Asboth2016short}.
In our case we can observe pumping along temporal or spatial directions
depending if we change $\varphi_{t}$ or $\varphi_{x}$. For pumping
in time (space), we obtain a clear illustration when we analyze transport
of Wannier states which are localized in a single site of temporal
(spatial) lattice and are not necessarily localized along spatial
(temporal) direction. To construct such Wannier states, we will choose
Floquet states which correspond to quasi-energy levels with different
temporal (spatial) index. Note that choosing Floquet states from a
given band, one obtains Wannier states which live in the corresponding
Hilbert subspace and which are uncoupled from states belonging to
any other bands.

\subsection{Thouless pumping in time}

We begin the analysis by considering time-pumping in a system of a
single spatial cell ($N=1$) which contains two sites originating
from the double-well structure of the spatial potential. The relevant
part of the quasienergy spectrum of $\hat{{\cal H}}$ is shown in
Fig.~\ref{fig:2D-pumping-time}(a). The figure displays the changes
of the quasienergy levels in the course of the adiabatic pumping in
the temporal dimension --- the temporal phase $\varphi_{t}$ is varied
while keeping $\varphi_{x}=0$. We interpret the obtained quasienergy
spectrum as follows. In the limit of an infinite spatial crystal,
the energy spectrum of $\hat{h}$ features series of bands separated
by large gaps. Because of the double-well structure of each cell of
the potential, each band consists of two sub-bands (a higher and a
lower one), separated by a small gap, of order $V_{{\rm L}}$ at $\varphi_{x}=0$.
Thus, each two consecutive levels in Fig.~\ref{fig:2D-pumping-time}(a)
correspond, respectively, to the higher and the lower sub-bands of
a certain spatial band. For example, the two topmost levels in Fig.~\ref{fig:2D-pumping-time}(a)
correspond to the sub-bands of the spatial band number 25 (counting
from the lowest); the spatial potential supports 28 clearly formed
bands in total with the chosen parameter values. From the time-crystalline
structure perspective, we associate the four topmost levels in Fig.~\ref{fig:2D-pumping-time}(a)
with the first temporal band, and the lower four levels with the second.
Note that it is natural to assign the lowest band number to the temporal
band whose quasienergy is largest: a particle confined in the temporal
lattice is parameterized by the effective mass which is negative (see
Appendix B) and its energy spectrum is thus bounded from above. To
distinguish the quasienergy levels constituting the first temporal
band, we introduce an index $\beta$. We assign the same index $\beta$
to all the levels corresponding to the same spatial band: we assign
$\beta=1$ to the two topmost levels in Fig.~\ref{fig:2D-pumping-time}(a)
and $\beta=2$ to the next two.

Analysis of the pumping process may be conveniently approached by
constructing the Wannier functions \citep{Resta1999,Soluyanov2011,Spaldin2012,Asboth2016short},
which are spatially localized superpositions of the relevant Floquet
modes: $w_{\alpha}(x,t)=\sum_{\beta=1}^{2}d_{\beta}^{(\alpha)}u_{\beta}(x,t)$,
where the coefficients $d_{\beta}^{(\alpha)}$ are found by diagonalizing
the position operator $\e^{2\i x/N}$ (see Appendix A). We will only
consider the quasienergy levels of the first temporal band. This is
justified by assuming that the gap between the first and the second
bands is large enough so that particles loaded into the first band
stay there throughout the pumping cycle. Furthermore, since we are
now considering pumping in the temporal direction only, we can restrict
our attention to a single site (out of two) of the spatial lattice.
To obtain Wannier functions that are localized in the same spatial
site, we have to mix the spatial energy levels corresponding either
to the higher spatial sub-bands or to the lower ones. We choose to
mix the levels of the higher spatial sub-bands by mixing the two modes
$\beta=1,2$ whose quasienergy levels are highlighted in Fig.~\ref{fig:2D-pumping-time}(a).
Mixing the other pair of levels (from the first temporal band) leads
to analogous results and corresponds to a particle occupying the other
site of the spatial lattice.

The obtained Wannier functions $w_{1}(x,t)$ and $w_{2}(x,t)$ are
shown in Fig.~\ref{fig:2D-pumping-time}(b) at $\varphi_{t}=0$ where
they are represented by black regions. In the present case we consider
only one spatial cell; the two sites of this cell span the regions
$x\in[0,\pi/4)\cup[3\pi/4,\pi)$ and $x\in[\pi/4,3\pi/4)$ --- note
that we assume periodic boundary conditions in space. The sites are
separated by white gaps in Fig.~\ref{fig:2D-pumping-time}(b). Each
of the spatial sites contains $s=2$ temporal cells, which we will
number with the index $\gamma=1,2$ and which are indicated by shaded
green and pink areas.\footnote{We reiterate the meaning of the indices $\alpha$, $\beta$, $\gamma$
to prevent confusion: index $\beta$ numbers the Floquet modes $u_{\beta}$,
while $\alpha$ numbers the Wannier functions $w_{\alpha}$ (which
are superpositions of the chosen Floquet modes). Independently, index
$\gamma$ numbers the cells of the temporal lattice; a Wannier function
$w_{\alpha}$ may in principle occupy any temporal cell.} We adopt the convention that the region of time-space which is occupied
by $w_{1}$ at $\varphi_{t}=0$ belongs to the first temporal cell
{[}$\gamma=1$, green shading in Fig.~\ref{fig:2D-pumping-time}(b){]},
while the region occupied by $w_{2}$ belongs to the second temporal
cell {[}$\gamma=2$, pink shading in Fig.~\ref{fig:2D-pumping-time}(b){]}.
Note that at a different value of $\varphi_{t}$, the state $w_{1}$
may spread over both temporal cells or even transition to cell $\gamma=2$,
and similarly for $w_{2}$. As illustrated in Fig.~\ref{fig:2D-pumping-time}(b),
the Wannier functions cycle in time between the two turning points
of the spatial site they are confined to, akin to classical pendula.

In space crystals we are interested in periodic distribution of particles
in space at a fixed moment of time (i.e.~the moment of the detection).
Switching from space to time crystals, the roles of space and time
are exchanged. That is, we fix position in space and ask if the probability
for the detection of particles at this fixed space-point changes periodically
in time \citep{SachaTC2020}. To understand the emergence of a time-crystalline
structure in the system analyzed here, let us consider placing a detector
close to the left (say) classical turning point $x_{0}$ of the spatial
lattice site under consideration. We take $x_{0}=0.3\pi$ for the
chosen energy regime, as indicated in Fig.~\ref{fig:2D-pumping-time}(b).
As shown in Fig.~\ref{fig:2D-pumping-time}(c) depicting the time-periodic
Wannier functions $w_{\alpha}(x_{0},t)$ at $\varphi_{t}=0$, in the
time intervals $(\Omega t\bmod2\pi)\in[0,\pi)$ the detector will
most likely be registering the particle whose wave function is $w_{1}(x_{0},t)$.
Meanwhile, in the intervals $(\Omega t\bmod2\pi)\in[\pi,2\pi)$ the
detector will most likely be registering the particle whose wave function
is $w_{2}(x_{0},t)$. The two time intervals can be considered to
divide the time axis into cells, allowing one to introduce the notion
of a time-crystalline structure. In the present case, the temporal
dimension of the crystalline structure is $2\pi/\Omega$, and the
crystal is periodic in this dimension, i.e.~periodic boundary conditions
in time are imposed. The lower panel of Fig.~\ref{fig:2D-pumping-time}(c)
demonstrates additionally that at a different value of the phase,
$\varphi_{t}=\pi/2$, the Wannier functions are shifted and they are
delocalized over both cells of the temporal lattice.

\begin{figure*}
\begin{centering}
\includegraphics{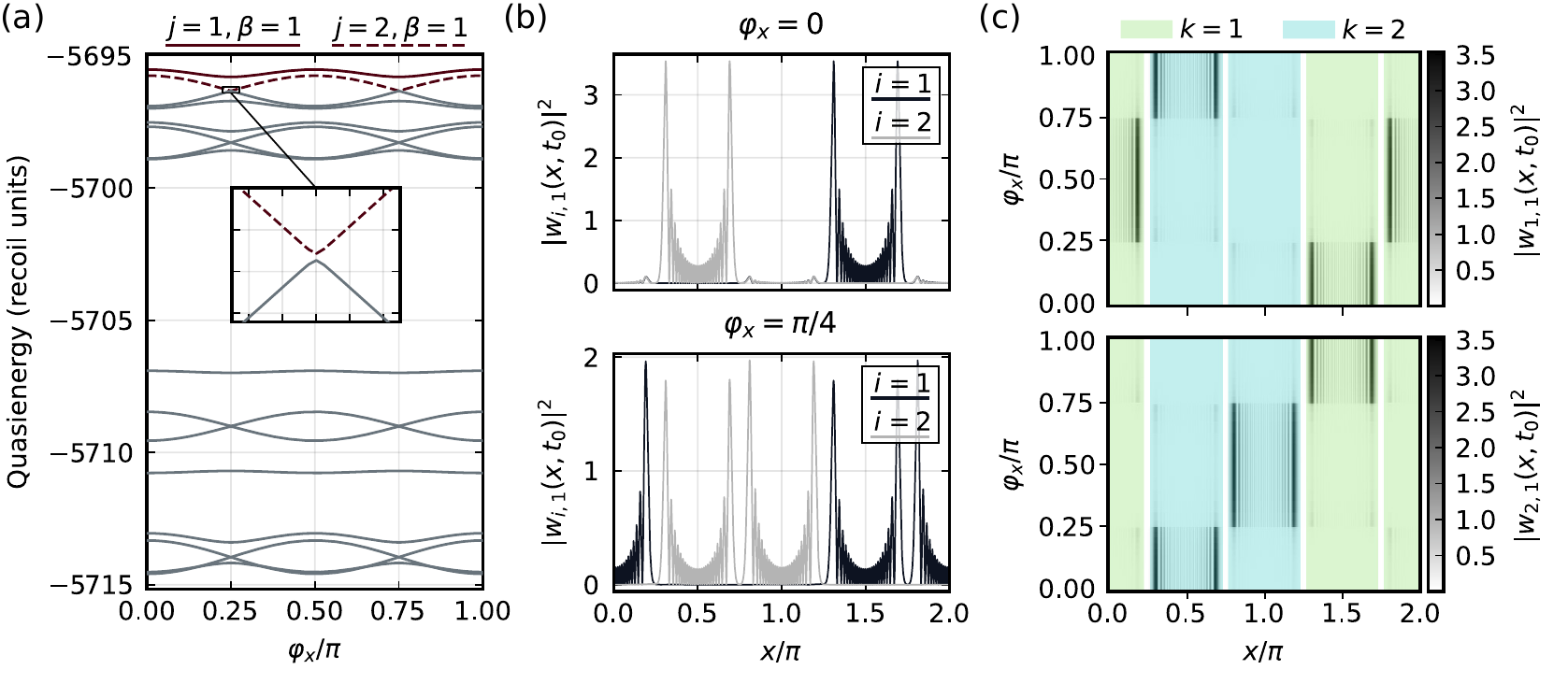}
\par\end{centering}
\caption{\label{fig:2D-pumping-space}Spatial adiabatic pumping in a 2D time-space
crystal with $s=2$ temporal cells and $N=2$ spatial cells.. The
same values of parameters were used as in Fig.~\ref{fig:2D-pumping-time}
except that $\varphi_{t}=0$, while $\varphi_{x}$ is varied. (a)
Quasienergy levels $\varepsilon_{j,\beta}$ of the Floquet Hamiltonian
$\hat{{\cal H}}$ versus the adiabatic phase $\varphi_{x}$. (b) Wannier
functions $|w_{i,1}(x,t_{0})|^{2}$ (where $t_{0}=\pi/2\Omega$) at
$\varphi_{x}=0$ and $\varphi_{x}=\pi/4$. (c) Changes of the Wannier
functions $|w_{i,1}(x,t_{0})|^{2}$ as $\varphi_{x}$ scans across
a complete cycle of length $\pi$. The cyan and green shaded areas
indicate the extent of the two spatial lattice cells ($k=1,2$), each
consisting of two sites separated by unshaded regions.}
\end{figure*}
Having introduced the concept of a crystalline structure in time,
we now turn to the Thouless pumping in the temporal dimension. To
this end, we calculate the Wannier functions repeatedly as the phase
$\varphi_{t}$ is varied and produce the plots of $w_{1}(x_{0},t)$
and $w_{2}(x_{0},t)$ for each value of $\varphi_{t}$, as shown in
Fig.~\ref{fig:2D-pumping-time}(d). We can clearly see that $w_{1}$
is being pumped from cell $\gamma=1$ to cell $\gamma=2$ as the temporal
phase $\varphi_{t}$ is varied from 0 to $2\pi$. At the same time,
state $w_{2}$ adiabatically transitions from cell 2 to cell 1. To
observe the pumping experimentally, one has to prepare a particle
in, e.g., the state $w_{1}(x,t)$ and place a detector at $x_{0}$.
Initially, the detector will most probably be detecting the particle
in the time intervals $(\Omega t\bmod2\pi)\in[0,\pi)$, whereas after
a pumping cycle is complete the detector will be clicking in the time
intervals $(\Omega t\bmod2\pi)\in[\pi,2\pi)$. We note in passing
that one can ``invert'' the pumping direction by letting $\varphi_{t}$
vary from 0 to $-2\pi$, just as is possible in the case of the adiabatic
pumping in real space.

\subsection{Thouless pumping in space}

Now let us analyze spatial-only pumping in a 2D time-space crystalline
structure consisting of $N=2$ spatial cells (and $s=2$ temporal
cells) with periodic boundary conditions. Doubling the number of spatial
cells leads to a twice greater number of Floquet quasienergy levels
compared to the case of $N=1$, as shown in Fig.~\ref{fig:2D-pumping-space}(a).
The physical origin of the levels is the same as in the preceding
discussion, and it is now immediately apparent which levels arise
from the higher and the lower spatial sub-bands. The gap between the
sub-bands is the smallest (but non-vanishing) at $\varphi_{x}=\pi/4$
and $\varphi_{x}=3\pi/4$ since all wells of the spatial potential
are of equal depth at these phases. Once the quasienergy spectrum
is obtained, we again switch to the Wannier representation, this time
introducing an additional spatial index $j$ to number the Floquet
modes and $i$ to number the Wannier functions: $w_{i,\alpha=1}(x,t)=\sum_{j=1}^{2}d_{j}^{(i)}u_{j,\beta=1}(x,t)$.
Since we are interested in the spatial pumping, we mix the Floquet
modes bearing the same temporal index $\beta=1$ so that the obtained
Wannier functions come out delocalized over the entire temporal lattice
structure, simplifying the analysis. The delocalization in the temporal
dimension and localization in spatial dimension means physically that
the particle, being confined to a single spatial site, can be detected
with equal probabilities at both turning points, this being true at
all times. This contrasts the situation in Sec.~III A, where, at
any given time, a particle could be detected at one turning point
with a higher probability than at the other, and based on this probability
we could speak of the particle occupying a specific temporal cell.
Note that the eigenstates of $\e^{2\i x/N}$ are always strongly localized
in certain sites of the spatial lattice, and so are the Wannier functions
$w_{i,\alpha}(x,t)$, at all $t$. The latter is only violated at
values of $\varphi_{x}$ close to $(2n+1)\pi/4$, $n\in\mathbb{Z}$,
when the depths of all the wells of the potential become equal, leading
to the Wannier functions spreading over two adjacent sites. The temporal
dependence of $w_{i,\alpha}(x,t)$, on the other hand, is dictated
by the temporal dependencies of the modes $u_{j,\beta}(x,t)$ that
are being mixed.

The relevant quasienergy levels are highlighted in Fig.~\ref{fig:2D-pumping-space}(a).
Their indices are $(j=1,\beta=1)$ and $(j=2,\beta=1)$ corresponding
to them occupying two different spatial cells. To analyze the pumping,
we study the states $w_{1,1}(x,t_{0})$ and $w_{2,1}(x,t_{0})$ at
a fixed detection moment $t_{0}=\pi/2\Omega$. Figure \ref{fig:2D-pumping-space}(b)
illustrates that, at $\varphi_{x}=0$, each of these states is localized
in a single site of the spatial lattice, while at $\varphi_{x}=\pi/4$
they occupy two sites in the process. We number the states and the
cells of the spatial lattice such that $w_{1,1}$ occupies spatial
cell $k=1$, while $w_{2,1}$ occupies spatial cell $k=2$ at the
beginning of the pumping cycle (at $\varphi_{x}=0$), as shown in
Fig.~\ref{fig:2D-pumping-space}(c). In the figure, the cyan and
green shaded areas indicate the spatial extent of the spatial lattice
cells, with the sites of the cells separated by unshaded gaps corresponding
to the positions of the barriers of the spatial potential. In the
end of the cycle, the Wannier states end up in a cell different from
the starting one, confirming that pumping does take place. It is apparent
that the states remain almost insensitive to the change of the potential
and are transported to a neighboring site abruptly. However, the lower
panel of Fig.~\ref{fig:2D-pumping-space}(b) demonstrates that the
transfer does not happen instantaneously, but rather proceeds via
a stage when the Wannier states occupy both sites.

We remark that constructing the Wannier functions using the Floquet
modes corresponding to the top third and fourth levels in Fig.~\ref{fig:2D-pumping-space}(a)
leads to pumping in the opposite direction around the circular $x$-axis
(not shown). This is to be expected since those energy levels correspond
to the lower spatial sub-bands, while the above results concern pumping
in the higher sub-bands \citep{Asboth2016short}.

\subsection{2D Thouless pumping}

\begin{figure*}
\begin{centering}
\includegraphics{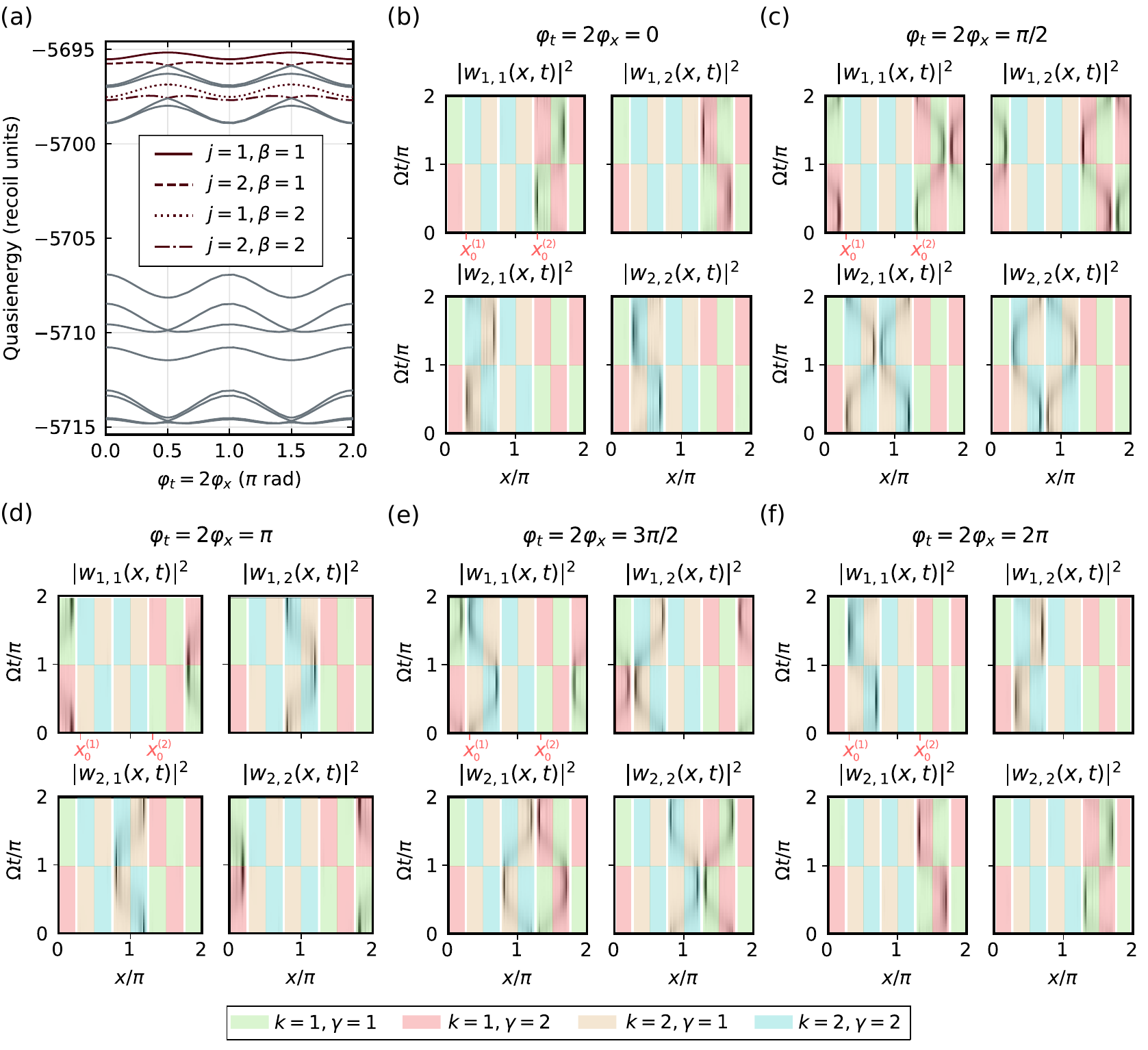}
\par\end{centering}
\caption{\label{fig:2D-pumping-time-space}Simultaneous temporal and spatial
adiabatic pumping in a 2D time-space crystal. The same values of parameters
were used as in Fig.~\ref{fig:2D-pumping-time} except that both
$\varphi_{t}$ and $\varphi_{x}$ are varied. (a) Quasienergy levels
$\varepsilon_{j,\beta}$ of the Floquet Hamiltonian $\hat{{\cal H}}$
versus the adiabatic phase $\varphi_{t}=2\varphi_{x}$. (b)--(f)
The Wannier functions at $\varphi_{t}=2\varphi_{x}=0,\pi/2,\pi,3\pi/2,2\pi$.
The probability densities $|w_{i,\alpha}(x,t)|^{2}$ are represented
by black regions, while the shaded areas indicate the extent of the
spatial ($k=1,2$) and temporal ($\gamma=1,2$) cells. In top-left
panels in (b)--(f), $x_{0}^{(1)}$ and $x_{0}^{(2)}$ indicate the
locations of two detectors, see text.}
\end{figure*}
We are now ready to discuss the simultaneous temporal and spatial
adiabatic pumping, demonstrated in Fig.~\ref{fig:2D-pumping-time-space}
for the case $N=s=2$. The adiabatic phases are varied along the trajectory
$\varphi_{t}=2\varphi_{x}$ from $\varphi_{t}=0$ to $\varphi_{t}=2\pi$
so that a complete pumping cycle is performed both in the temporal
and in the spatial dimensions. The obtained Floquet quasienergy spectrum
is shown in Fig.~\ref{fig:2D-pumping-time-space}(a), where the legend
indicates the quasienergy levels corresponding to the modes that we
mix when constructing the Wannier states. The relevant modes are those
of the first temporal band, among which we select those corresponding
to the higher spatial sub-bands. As discussed in Sec.~III B, this
selection allows us to focus on the Wannier states that are transported
in space to the right during the pumping process. The remaining four
levels of the first temporal band constitute the Wannier functions
that are being pumped to the left in space. The constructed states
are expressed as $w_{i,\alpha}(x,t)=\sum_{j,\beta}d_{j,\beta}^{(i,\alpha)}u_{j,\beta}(x,t)$
and are obtained as above, by diagonalizing the position operator
$\e^{2\i x/N}$. Contrary to the preceding analysis, we no longer
restrict our attention to a certain position $x_{0}$ or a certain
detection time $t_{0}$, but rather study the two-dimensional maps
of the Wannier functions $w_{i,\alpha}(x,t)$. The four Wannier states
are shown in Figs.~\ref{fig:2D-pumping-time-space}(b)--(f) at various
values of the adiabatic phases, with the shaded areas dividing the
whole time-space into temporal and spatial cells. The states and the
cells are numbered so that initially (at $\varphi_{t}=2\varphi_{x}=0$)
the Wannier state indices $(i,\alpha)$ coincide with the spatial
and temporal cell numbers $(k,\gamma)$. In each of the panels (b)--(f),
four sites are left unoccupied by any Wannier functions --- these
would be occupied by the Wannier functions constructed using the states
of the lower spatial sub-bands, i.e.~the Floquet modes corresponding
to the top four unhighlighted quasienergy levels in Fig.~\ref{fig:2D-pumping-time-space}(a).
We note in passing that the two Wannier functions constructed in Sec.~III
B using the states corresponding to only the two upper quasienergy
levels appear as the sums $w_{1,1}+w_{1,2}$ and $w_{2,1}+w_{2,2}$,
where $w_{i,\alpha}$ are the functions displayed in Figs.~\ref{fig:2D-pumping-time-space}(b)--(f).
Such sums exhibit spatial, but not temporal localization.

Turning to the pumping process, in Fig.~\ref{fig:2D-pumping-time-space}(c)
we see the Wannier states are transported to the neighboring spatial
site (to the right) as a result of the spatial pumping, while the
temporal pumping causes the states to slide down the temporal axis.
At $\varphi_{t}=2\varphi_{x}=\pi$ {[}see Fig.~\ref{fig:2D-pumping-time-space}(d){]},
the spatial transition is complete, whereas the time dependence of
the functions is such that the functions occupy both temporal cells.
Next, at $\varphi_{t}=2\varphi_{x}=3\pi/2$ {[}see Fig.~\ref{fig:2D-pumping-time-space}(e){]},
the states are shown in the middle of the second spatial transition,
which is completed at $\varphi_{t}=2\varphi_{x}=2\pi$ {[}see Fig.~\ref{fig:2D-pumping-time-space}(f){]}.
Comparing Figs.~\ref{fig:2D-pumping-time-space}(b) and \ref{fig:2D-pumping-time-space}(f),
it is apparent that as a result of the pumping each state $w_{i,\alpha}$
has transitioned from cell $(i,\alpha)$ to cell $(i+1\mod N,\ \alpha+1\mod s)$.

Let us now give an interpretation of these results. Consider two detectors,
one placed at $x_{0}^{(1)}=0.3\pi$ and the other one at $x_{0}^{(2)}=1.3\pi$,
and a particle loaded initially into the system in the state $w_{1,1}$.
At $\varphi_{t}=2\varphi_{x}=0$, most probably a detector placed
at $x_{0}^{(2)}$ will be detecting the particle in the time intervals
$(\Omega t\bmod2\pi)\in[0,\pi)$, corresponding to the particle occupying
the first spatial and the first temporal cells {[}see Fig.~\ref{fig:2D-pumping-time-space}(b){]}.
In the end of the pumping cycle ($\varphi_{t}=2\varphi_{x}=2\pi$),
the particle will most probably appear in the time intervals $(\Omega t\bmod2\pi)\in[\pi,2\pi)$
on a detector placed at $x_{0}^{(1)}$, corresponding to the particle
occupying the second spatial and the second temporal cells.

\section{Conclusions}

Summarizing our work, we have shown that the quasienergy spectrum
of a resonantly driven optical lattice may be interpreted as that
of a crystal-like structure with the time playing the role of an additional
coordinate. Using this analogy, we studied adiabatic variation of
the driving protocol and demonstrated that it leads to a change of
system dynamics that is a manifestation of the Thouless pumping in
the temporal dimension. Finally, we have illustrated simultaneous
adiabatic pumping in both spatial and temporal directions.
\begin{acknowledgments}
We would like to thank Alexandre Dauphin for a stimulating discussion on the Thouless pumping in time. 
Support of the National Science Centre, Poland, via Project No.~2018/31/B/ST2/00349 (C.-H.\,F.) is acknowledged. 
This research was also funded in part by the National Science Centre, Poland, Project No.~2021/42/A/ST2/00017 (K.\,S.). 
For the purpose of Open Access, the author has applied a CC-BY public copyright licence to any Author Accepted 
Manuscript (AAM) version arising from this submission.

\end{acknowledgments}

\section*{Appendix A: Diagonalization of Floquet Hamiltonian}

\setcounter{equation}{0}
\renewcommand{\theequation}{A\arabic{equation}}

In this section, we discuss the diagonalization of the Floquet Hamiltonian
\begin{equation}
\hat{{\cal H}}=\hat{h}-\i\frac{\partial}{\partial t}+\xi_{{\rm S}}+\xi_{{\rm L}},\label{eq:floqop}
\end{equation}
where $\hat{h}$ is defined in Eq.~(\ref{eq:h}), while $\xi_{{\rm S}}$
and $\xi_{{\rm L}}$ are given in Eqs.~(\ref{eq:perturb}).

In order to solve the eigenvalue problem
\begin{equation}
\hat{{\cal H}}u_{n}(x,t)=\varepsilon_{n}u_{n}(x,t),
\end{equation}
we first numerically obtain the eigenstates of $\hat{h}$ in the basis
of plane waves $\braket xj=\e^{\i\frac{2j}{N}x}/\sqrt{N\pi}$ orthonoromal
on $x\in[0,N\pi)$, where $N$ is the number of spatial cells. The
sought eigenstates fulfill 
\begin{equation}
\hat{h}\psi_{m}(x)=\epsilon_{m}\psi_{m}(x),
\end{equation}
are written as 
\begin{equation}
\psi_{m}(x)=\frac{1}{\sqrt{N\pi}}\sum_{j=-\infty}^{\infty}c_{j}^{(m)}\,\e^{\i\frac{2j}{N}x},
\end{equation}
and the coefficients $c_{j}^{(m)}$ that express the solution in the
$m$th energy band are obtained by diagonalizing the matrix
\begin{equation}
\begin{split}\bra{j'}h\ket j & =\left[\left(\frac{2j}{N}\right)^{2}+\frac{V_{{\rm S}}+V_{{\rm L}}}{2}\right]\delta_{j',j}\\
 & +\frac{V_{{\rm S}}}{4}(\delta_{j',j+2N}+\delta_{j',j-2N})\\
 & +\frac{V_{{\rm L}}}{4}(\e^{2\i\varphi_{x}}\delta_{j',j+N}+\e^{-2\i\varphi_{x}}\delta_{j',j-N}).
\end{split}
\end{equation}
Once we have the eigenstates of the unperturbed Hamiltonian, an additional
transformation to the rotating frame provides a suitable basis consisting
of functions 
\begin{equation}
\psi'_{m}(x,t)=\e^{-\i\nu(m)\omega t/s}\psi_{m}(x).\label{eq:psi'}
\end{equation}
Here, the function $\nu(m)=\lceil m/2N\rceil$ (where $\lceil\cdots\rceil$
is the ceiling operation) transforms the level numbers $m=1,2,3,4,5,6,\ldots$
into band indices
\begin{equation}
\nu=\underset{2N\text{ times}}{\underbrace{1,\ldots,1}},\underset{2N\text{ times}}{\underbrace{2,\ldots,2}},\underset{2N\text{ times}}{\underbrace{3,\ldots,3}},\underset{2N\text{ times}}{\underbrace{4,\ldots,4}},\ldots
\end{equation}
Note that this labeling is correct when the sites of a potential cell
are not too asymmetric for all $\varphi_{x}$ --- otherwise one has
to examine how to properly label the unperturbed eigenstates so that
the unitary transformation (\ref{eq:psi'}) corresponds to the canonical
transformation to the moving frame (see Appendix B). In practice,
we have to keep $V_{{\rm L}}$ small enough so that the difference
of depths of the potential wells in each cell is always smaller than
the gaps between the energy bands.

We now calculate the matrix elements of $\hat{{\cal H}}$. For the
diagonal part, we have 
\begin{equation}
\bra{\psi'_{m'}}\left(\hat{h}-\i\frac{\partial}{\partial t}\right)\ket{\psi'_{m}}=\left[\epsilon_{m}-\frac{\nu(m)\omega}{s}\right]\delta_{m',m}.\label{eq:diag}
\end{equation}

For the long perturbation, we obtain
\begin{equation}
\begin{split} & \bra{\psi'_{m'}}\xi_{{\rm L}}\ket{\psi'_{m}}=\lambda_{{\rm L}}\\
 & \qquad\times\cos(\omega t+\varphi_{t})\,\e^{-\i(\omega/s)[\nu(m)-\nu(m')]t}\\
 & \qquad\times\sum_{j,j'}c_{j'}^{(m')*}c_{j}^{(m)}\int_{0}^{\pi}\frac{\d x}{\pi}{\cal Q}_{{\rm L}}(x)\,\e^{\i(2j-2j')x},
\end{split}
\end{equation}
where ${\cal Q}_{{\rm L}}$ is the spatial part of $\xi_{{\rm L}}$,
see Eqs.~(\ref{eq:perturb}). Applying the secular approximation,
we replace
\begin{equation}
\cos(\omega t+\varphi_{t})\,\e^{-\i(\omega/s)[\nu(m)-\nu(m')]t}
\end{equation}
with its time-independent contribution
\begin{equation}
\tfrac{1}{2}(\e^{\i\varphi_{t}}\delta_{\nu'+s,\nu}+\e^{-\i\varphi_{t}}\delta_{\nu'-s,\nu}),
\end{equation}
where $\nu'\equiv\nu(m')$. With our choice ${\cal Q}_{{\rm L}}(x)=\cos^{2}(2x)$,
we finally obtain 
\begin{multline}
\bra{\psi'_{m'}}\xi_{{\rm L}}\ket{\psi'_{m}}=\frac{\lambda_{{\rm L}}}{2}(\e^{\i\varphi_{t}}\delta_{\nu'+s,\nu}+\e^{-\i\varphi_{t}}\delta_{\nu'-s,\nu})\\
\times\frac{1}{4}\sum_{j=-\infty}^{\infty}c_{j}^{(m)}(2c_{j}^{(m')*}+c_{j+2}^{(m')*}+c_{j-2}^{(m')*}).\label{eq:xi_L}
\end{multline}

Similarly, using ${\cal Q}_{{\rm S}}(x)=\sin^{2}(2x)$, the matrix
elements of the short perturbation follow as

\begin{multline}
\bra{\psi'_{m'}}\xi_{{\rm S}}\ket{\psi'_{m}}=\frac{\lambda_{{\rm S}}}{2}(\e^{\i\varphi_{t}}\delta_{\nu'+2s,\nu}+\e^{-\i\varphi_{t}}\delta_{\nu'-2s,\nu})\\
\times\frac{1}{4}\sum_{j=-\infty}^{\infty}c_{j}^{(m)}(2c_{j}^{(m')*}-c_{j+2}^{(m')*}-c_{j-2}^{(m')*}).\label{eq:xi_S}
\end{multline}

Once the eigenfunctions $u_{n}(x,t)=\sum_{m}b_{m}^{(n)}\psi'_{m}(x,t)$
are found, the Wannier functions are constructed by diagonalizing
the periodic position operator $\e^{\i\frac{2}{N}x}$ \citep{Asboth2016short,Aligia1999},
whose matrix elements result as
\begin{equation}
\begin{split}\bra{u_{n'}}\e^{\i\frac{2}{N}x}\ket{u_{n}} & =\sum_{j=-\infty}^{\infty}\sum_{m,m'}b_{m'}^{(n')*}b_{m}^{(n)}c_{j+1}^{(m')*}c_{j}^{(m)}\\
 & \qquad\times\e^{\i[\nu(m')-\nu(m)]\omega t/s}.
\end{split}
\end{equation}
The diagonalization is performed at a chosen time moment $t$. The
obtained coefficients $d_{n}^{(k)}$ are used to express the Wannier
states as 
\begin{equation}
w_{k}(x,t)=\sum_{n}d_{n}^{(k)}u_{n}(x,t).
\end{equation}
Note that the position operator is constructed using only the relevant
subspace of eigenvectors $\ket{u_{n}}$ that we select based on our
interpretation of the quasienergy spectrum, as discussed in the main
text. Subsequent renumbering of these eigenvectors and the Wannier
states using a pair of indices $(i,\alpha)$ is likewise conventional.

All calculations have been performed using a number of software packages
\citep{JuliaDiffEq,JuliaDiffEq2,JuliaDiffEqKahanLi,JuliaDiffEqMcAte,JuliaOptim}
written in the Julia programming language \citep{Julia}.

\section*{Appendix B: Quasiclassical analysis of the temporal Thouless pumping}

\setcounter{equation}{0}
\setcounter{figure}{0}
\renewcommand{\thefigure}{B\arabic{figure}}
\renewcommand{\theequation}{B\arabic{equation}}

It is instructive to perform an analysis of a one-dimensional time
crystal, whereby a particle is confined to a single potential well
and the spatial periodicity plays no role. Rather than using the Floquet
theory, we can consider our model Hamiltonian
\begin{equation}
H(p_{x},x,t)=h(p_{x},x|\varphi_{x})+\xi_{{\rm S}}(x,t)+\xi_{{\rm L}}(x,t|\varphi_{t})
\end{equation}
as a classical entity and use the action--angle representation of
the unperturbed Hamiltonian $h(I)$, with the action $I$ and angle
$\theta\in[0,2\pi)$ constituting a pair of canonical variables \citep{Lichtenberg1992}.
Specifically, the action variable is defined for a one-dimensional
time-independent Hamiltonian {[}$h(p_{x},x|\varphi_{x})$ in our case{]}
as the integral of momentum along a periodic orbit 
\begin{equation}
I=\frac{1}{2\pi}\oint p_{x}\,\d x.
\end{equation}
The angle $\theta$ is the position variable of a particle on a periodic
trajectory that changes uniformly in time, $\theta(t)=\Omega t+\theta(0)$,
where the frequency of the periodic motion $\Omega=\partial h(I)/\partial I$.
The periodic motion of a classical particle confined to a single lattice
site of the potential in $h$ is represented in the $(I,\theta)$
phase space by straight lines $I(\theta)={\rm const}$ when no time-dependent
perturbation is present. The perturbation causes formation of resonant
islands that contain closed orbits in the vicinity of the resonant
value of action $I_{s}$ such that $\Omega=[\partial h(I)/\partial I]\big|_{I_{s}}=\omega/s$
where $s$ in an integer \citep{Buchleitner2002}. Analysis of the
motion may be simplified by transitioning to the frame moving along
the resonant orbit and applying the secular approximation, whereby
the oscillatory terms of the Hamiltonian are dropped. We give all
the details of this calculation below.

Setting $\varphi_{x}=0$ and performing a transformation to the action--angle
variables, we obtain 
\begin{equation}
\begin{split}H(I,\theta,t) & =h(I)\\
 & +\lambda_{{\rm S}}\cos(2\omega t){\cal Q}_{{\rm S}}(I,\theta)\\
 & +\lambda_{{\rm L}}\cos(\omega t+\varphi_{t}){\cal Q}_{{\rm L}}(I,\theta),
\end{split}
\label{eq:H-1}
\end{equation}
where ${\cal Q}_{{\rm S}}(x)=\sin^{2}(2x)$ and ${\cal Q}_{{\rm L}}(x)=\cos^{2}(2x)$
--- we have chosen different functions for ${\cal Q}_{{\rm S}}(x)$
and ${\cal Q}_{{\rm L}}(x)$ but one can also choose $\sin(2x)$ {[}or
$\cos(2x)${]} for both of them. The spatial Hamiltonian $h(I)$ does
not depend on $\theta$, and the $\theta$-dependencies of the perturbations
can be represented as Fourier series
\begin{equation}
{\cal Q}_{{\rm S}/{\rm L}}(I,\theta)=\sum_{m=-\infty}^{\infty}{\cal Q}_{{\rm S}/{\rm L}}^{(m)}(I)\,\e^{\i m\theta}.
\end{equation}

We choose our working point in the vicinity of a certain resonant
trajectory corresponding to a given value of the action $I_{s}$ and
the corresponding intrinsic frequency $\Omega$. Switching to the
rotating frame according to $\Theta=\theta-\Omega t$ and averaging
out the rapidly oscillating terms, we express the long perturbation
as
\begin{equation}
\xi_{{\rm L}}=\frac{1}{2}\lambda_{{\rm L}}\left[{\cal Q}_{{\rm L}}^{(s)}(I_{s})\,\e^{\i(\varphi_{t}-s\Theta)}+{\rm c.c.}\right].
\end{equation}
Writing the complex number ${\cal Q}_{{\rm L}}^{(s)}(I_{s})$ in the
polar form 
\begin{equation}
{\cal Q}_{{\rm L}}^{(s)}(I_{s})\equiv A_{{\rm L}}\,\e^{\i\chi_{{\rm L}}},
\end{equation}
we cast this result into
\begin{equation}
\xi_{{\rm L}}=\lambda_{{\rm L}}A_{{\rm L}}\cos(s\Theta-\chi_{{\rm L}}-\varphi_{t}).
\end{equation}

In complete analogy, the short perturbation is written as
\begin{equation}
\xi_{{\rm S}}=\lambda_{{\rm S}}A_{{\rm S}}\cos(2s\Theta-\chi_{{\rm S}}),
\end{equation}
where $A_{{\rm S}}\,\e^{\i\chi_{S}}$ is the polar form of ${\cal Q}_{S}^{(2s)}(I_{s})$.

We also expand $h(I)$ around $I_{s}$ to the second order in $I$,
\begin{equation}
h(I)=h(I_{s})+(I-I_{s})\Omega+\frac{1}{2}h''(I_{s})(I-I_{s})^{2}.
\end{equation}
Noting also that the time-dependent canonical transformation introduces
an additional term $-I\Omega$, we finally derive the effective Hamiltonian
\begin{equation}
\begin{split}H_{{\rm eff}} & =[h(I_{s})-\Omega I_{s}]+\frac{P^{2}}{2M}\\
 & +\lambda_{{\rm S}}A_{{\rm S}}\cos(2s\Theta-\chi_{{\rm S}})\\
 & +\lambda_{{\rm L}}A_{{\rm L}}\cos(s\Theta-\chi_{{\rm L}}-\varphi_{t}),
\end{split}
\label{eq:H_eff}
\end{equation}
with $P\equiv I-I_{s}$. Here we have also defined the ``effective
mass'' $M=1/h''(I_{s})$, which is negative. The last two lines of
Eq.~(\ref{eq:H_eff}) represent a periodic potential in the moving
frame. For $s=2$, we obtain a lattice of two elementary cells, each
consisting of two sites arising from the double-well structure.

\begin{figure}
\begin{centering}
\includegraphics{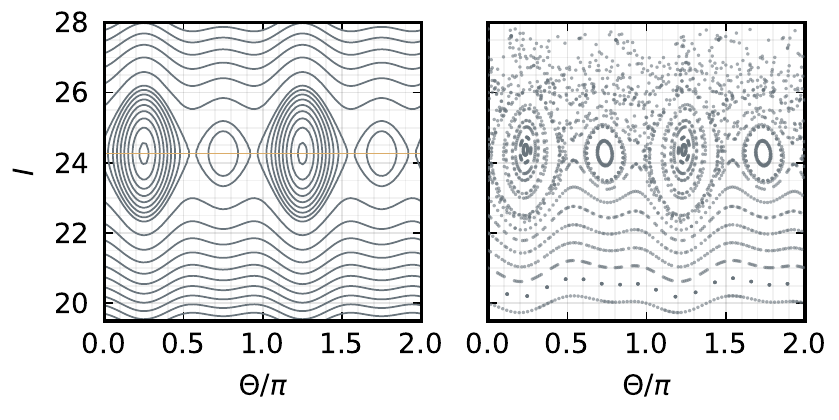}
\par\end{centering}
\caption{\label{fig:maps}Phase-space maps of the particle motion governed
by the secular effective Hamiltonian (\ref{eq:H_eff}) (left panel)
and the exact Hamiltonian (\ref{eq:H-1}) (right panel). The yellow
line in the left panel indicates the working point $I_{s}=24.3$.
The map in the right panel was generated by numerically integrating
the classical equations of motion resulting from the exact Hamiltonian
(\ref{eq:H-1}) and registering particle's position $(\theta,I)$
stroboscopically at the intervals of the lattice driving period. The
same values of parameters were used as in Fig.~\ref{fig:2D-pumping-time}.}
\end{figure}

In order to verify the validity of the secular approximation, we produce
a map of the motion of a particle in the $(I,\theta)$ phase-space
governed by the secular effective Hamiltonian (\ref{eq:H_eff}), as
shown in Fig.~\ref{fig:maps}, left panel. The right panel displays
the map obtained by integrating the exact equations of motion resulting
from the Hamiltonian (\ref{eq:H-1}) and by registering particle's
coordinates stroboscopically at the intervals of the lattice driving
period. Such a stroboscopic, rather than continuous, picture precisely
corresponds to the secular approximation, which only provides the
information on the dynamics averaged over the driving period. Comparing
the two plots in Fig.~\ref{fig:maps}, we conclude that the resonant
islands where the quantum states we are interested in will be located
(in the semiclassical sense) are well reproduced in the exact picture,
and hence that the secular approximation is valid for the considered
strength of the perturbation (i.e.~values of $\lambda_{{\rm S}}$
and $\lambda_{{\rm L}}$). 

We are now in position to quantize the classical effective Hamiltonian
(\ref{eq:H_eff}) by changing $\Theta\to\hat{\Theta}=\Theta$ and
$P\to\hat{P}=-\i\partial/\partial\Theta$:
\begin{equation}
\begin{split}\hat{H}_{{\rm eff}} & =[h(I_{s})-\Omega I_{s}]-\frac{1}{2M}\frac{\partial^{2}}{\partial\Theta^{2}}\\
 & +\lambda_{{\rm S}}A_{{\rm S}}\cos(2s\Theta-\chi_{{\rm S}})\\
 & +\lambda_{{\rm L}}A_{{\rm L}}\cos(s\Theta-\chi_{{\rm L}}-\varphi_{t}).
\end{split}
\label{eq:H_eff-q}
\end{equation}
This form of the Hamiltonian is particularly convenient since it features
an explicit expression for the temporal potential, which is not the
case for Eq.~(\ref{eq:H}). In complete analogy with conventional
space crystals, we may consider the limit of a large number of cells,
$s\gg1$ (while $\Theta\in[0,2\pi)$). In that case, the eigenstates
of Hamiltonian (\ref{eq:H_eff-q}) are Bloch waves given by $\psi_{n,{\cal K}}(\Theta)=\e^{\i{\cal K}\Theta}u_{n,{\cal K}}(\Theta)$,
where ${\cal K}$ is the time-quasimomentum and $u_{n,{\cal K}}(\Theta)=u_{n,{\cal K}}(\Theta+2\pi/s)$
are cell-periodic functions \citep{Zlabys2021}. Relevant for the
Thouless pumping, one may introduce the Berry curvature related to
the $n$th energy band as $\Omega_{n}(\varphi_{t},{\cal K})=\i(\braket{\partial_{\varphi_{t}}u_{n}}{\partial_{{\cal K}}u_{n}}-\braket{\partial_{{\cal K}}u_{n}}{\partial_{\varphi_{t}}u_{n}})$
and the corresponding first Chern number \citep{Xiao2010,Nakajima2016,Lohse2016pump}
\begin{equation}
\nu_{n}=\frac{1}{2\pi}\int_{{\rm BZ}}\d{\cal K}\intop_{0}^{2\pi}\d\varphi_{t}\,\Omega_{n}(\varphi_{t},{\cal K}),\label{eq:chern}
\end{equation}
where we integrate over a Brillouin zone in ${\cal K}$-space as the
phase $\varphi_{t}$ completes a pumping cycle.

Numerical diagonalization of (\ref{eq:H_eff-q}) is straightforward:
in order to solve the eigenvalue problem
\begin{equation}
\hat{H}_{{\rm eff}}\psi_{\beta}(\Theta)=E_{\beta}\psi_{\beta}(\Theta)
\end{equation}
under periodic boundary conditions, $\psi_{\beta}(\Theta)=\psi_{\beta}(\Theta+2\pi)$,
we expand the eigenfunctions in the basis of plane waves $\braket{\Theta}j=\e^{\i j\Theta}/\sqrt{2\pi},$
$j\in\mathbb{Z}$, which are orthonoromal on $\Theta\in[0,2\pi)$:
\begin{equation}
\psi_{\beta}(\Theta)=\frac{1}{\sqrt{2\pi}}\sum_{j=-\infty}^{\infty}c_{j}^{(\beta)}\,\e^{\i j\Theta}.
\end{equation}
This leads to the following matrix elements:
\begin{align}
\bra{j'}\hat{H}_{{\rm eff}}\ket j & =\left[h(I_{s})-\Omega I_{s}+\frac{j^{2}}{2M}\right]\delta_{j',j}\nonumber \\
 & +\frac{\lambda_{{\rm S}}A_{{\rm S}}}{2}(\delta_{j',j+2s}\,\e^{-\i\chi_{{\rm S}}}+\delta_{j',j-2s}\,\e^{\i\chi_{{\rm S}}})\\
 & +\frac{\lambda_{{\rm L}}A_{{\rm L}}}{2}(\delta_{j',j+s}\,\e^{-\i(\chi_{{\rm L}}+\varphi_{t})}+\delta_{j',j-s}\,\e^{\i(\chi_{{\rm L}}+\varphi_{t})}).\nonumber 
\end{align}

The final ingredient needed for the analysis is the Wannier functions,
which are localized superpositions of the stationary states of $\hat{H}_{{\rm eff}}$.
They may be found by diagonalizing the position operator. In the case
of a periodic system, we take this operator in the form $\e^{\i\Theta}$
\citep{Aligia1999,Asboth2016short}. Its matrix elements follow as
\begin{equation}
\bra{\psi_{\beta'}}\e^{\i\Theta}\ket{\psi_{\beta}}=\sum_{j=-\infty}^{\infty}c_{j+1}^{(\beta')*}c_{j}^{(\beta)}.
\end{equation}
Numerical diagonalization of this operator yields the coefficients
$d_{\beta}^{(\alpha)}$ that allow us to express the Wannier states
as 
\begin{equation}
w_{\alpha}(\Theta)=\sum_{\beta}d_{\beta}^{(\alpha)}\psi_{\beta}(\Theta).
\end{equation}

\begin{figure*}
\begin{centering}
\includegraphics{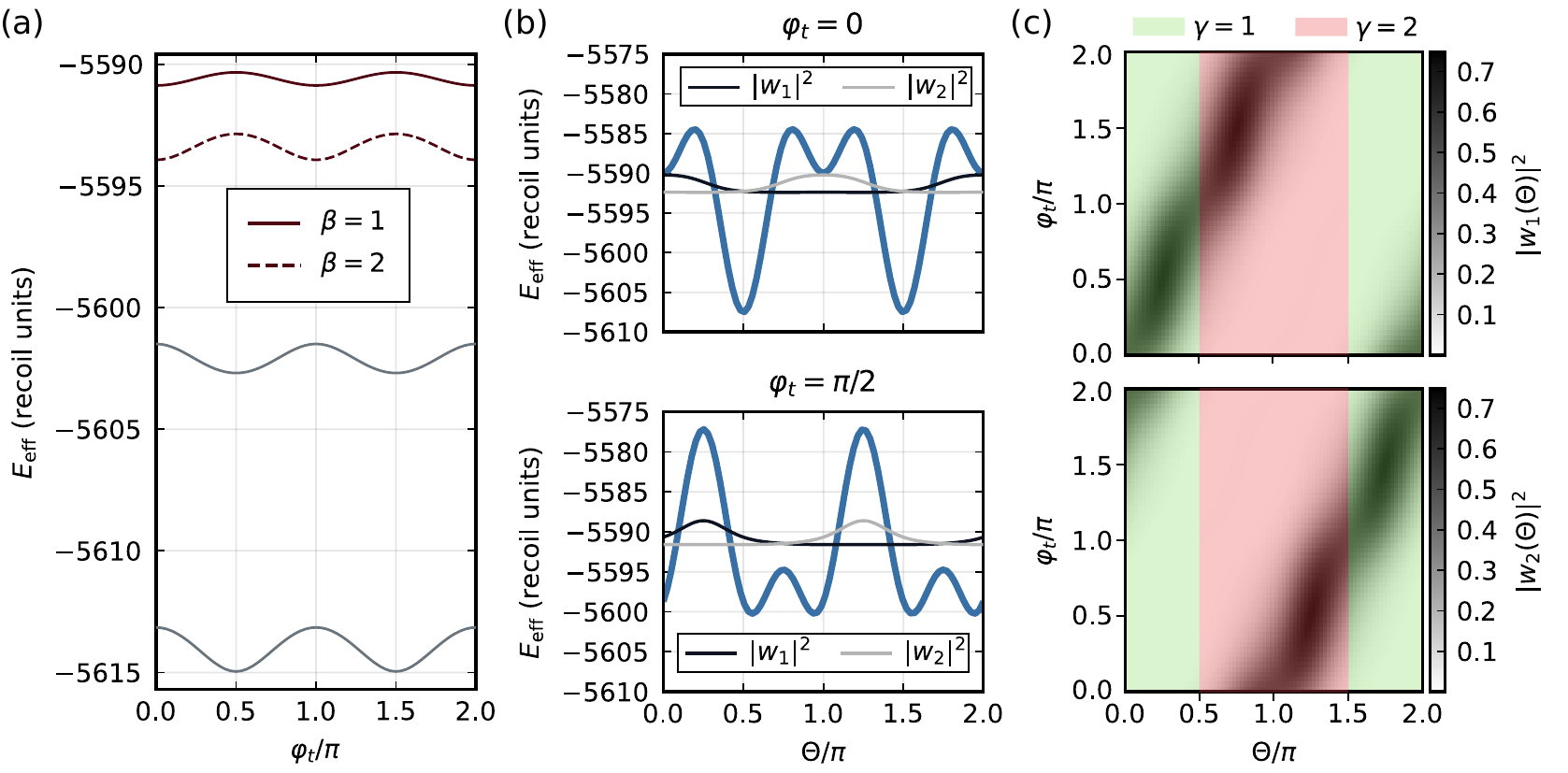}
\par\end{centering}
\caption{\label{fig:time-pumping}Adiabatic pumping in a two-cell time crystal
simulated based on the quasiclassical approach. The same values of
parameters were used as in Fig.~\ref{fig:2D-pumping-time}. (a) Energy
levels $E_{\beta}$ of quantized effective Hamiltonian (\ref{eq:H_eff-q})
versus the adiabatic phase $\varphi_{t}$. (b) Wannier functions $|w_{\alpha}(\Theta)|^{2}$
superimposed on the effective potential (blue curves) $U_{{\rm eff}}=H_{{\rm eff}}-P^{2}/2M$
at $\varphi_{t}=0$ and $\varphi_{t}=\pi/2$. The scale of the $y$-axis
relates to the potential. The Wannier functions are depicted in arbitrary
units; they are positioned such that the flat tails mark the corresponding
mean energies $\langle w_{\alpha}|\hat{H}_{{\rm eff}}|w_{\alpha}\rangle$.
(c) Changes of the Wannier functions $|w_{\alpha}(\Theta)|^{2}$ as
$\varphi_{t}$ scans across a complete cycle of length $2\pi$. The
shaded areas indicate the extent of the two temporal lattice cells
($\gamma=1,2$).}
\end{figure*}

Starting with the calculation of the eigenvalues of $\hat{H}_{{\rm eff}}$,
we look for the highest ones since in the case of negative mass the
energy spectrum is bounded from above. Four highest energy levels
calculated repeatedly as the adiabatic phase $\varphi_{t}$ is varied
are shown in Fig.~\ref{fig:time-pumping}(a). We interpret the two
highest levels ($\beta=1,2$) as belonging to the first energy band
and the next two as constituting the second band, with the bands being
separated by a gap. We note that the spectrum is similar to the one
in Fig.~\ref{fig:2D-pumping-time}(a) in the main text, except that
there we consider two spatial sites instead of only one, leading to
twice greater number of (quasi)energy levels. This similarity supports
the validity of the quasiclassical analysis and quantization of the
effective classical Hamiltonian (\ref{eq:H_eff}) in particular.

We use the eigenstates $\psi_{1}$ and $\psi_{2}$ corresponding to
the energy levels highlighted in Fig.~\ref{fig:time-pumping}(a)
to construct the Wannier functions $w_{1}$ and $w_{2}$, shown at
$\varphi_{t}=0$ and $\varphi_{t}=\pi/2$ in Fig.~\ref{fig:time-pumping}(b).
These functions are vertically positioned such that the flat tails
mark the corresponding mean energies $\langle w_{\alpha}|\hat{H}_{{\rm eff}}|w_{\alpha}\rangle$.
As we can see, the Wannier states are localized in the sites of the
effective potential $U_{{\rm eff}}(\Theta)=H_{{\rm eff}}-P^{2}/2M$
{[}blue curves in Fig.~\ref{fig:time-pumping}(b){]}, which may be
regarded as a crystalline structure for $s\gg1$. Switching back to
the lab frame and considering placing a detector at a fixed position
$\theta=\theta_{0}$, we recover the time periodicity of the potential
$U_{{\rm eff}}(\theta_{0}+\Omega t)$. Crucially, we can now interpret
the dynamics of the system in terms of the time-periodic Wannier states
$w_{\alpha}(\theta_{0}+\Omega t)$: their localization in $\Theta$-space
in the moving frame translates into localization in time in the lab
frame. Therefore, these states can be understood as being localized
in the cells of a time crystal. A stationary detector placed at $\theta_{0}$
in the lab frame will register periodic arrival of the two (for $s=2$)
Wannier states separated by the interval of $\pi/\Omega$, corresponding
to a detector scanning across $\Theta$ in the moving frame {[}cf.
Fig.~\ref{fig:time-pumping}(b){]}.

Now let us study the changes of the Wannier functions as $\varphi_{t}$
is varied adiabatically from 0 to $2\pi$. Figure \ref{fig:time-pumping}(b)
shows that at the beginning of the cycle ($\varphi_{t}=0$), state
$w_{1}$ occupies the region $\Theta\in[-\pi/2,\pi/2)$, which, by
convention, we will refer to as the first cell ($\gamma=1$) of the
temporal lattice. Similarly, $w_{2}$ occupies the region $\Theta\in[\pi/2,3\pi/2)$,
which we will call the second temporal cell ($\gamma=2$). Further
change of the two states is presented in Fig.~\ref{fig:time-pumping}(c).
It is apparent that the probability densities $|w_{\alpha}|^{2}$
shift as the phase is increased. By the end of the cycle, the Wannier
functions are seen to have shifted by $\pi$, with the state $w_{1}$
now occupying the second temporal cell, and $w_{2}$ occupying the
first. Note that the pumping direction appears to be reversed in Fig.~\ref{fig:time-pumping}(c)
compared to Fig.~\ref{fig:2D-pumping-time}(d) in the main text because
of the minus sign in the transformation $\Theta=\theta-\omega t/s$.
Since the number of particles pumped through a cross-section of the
lattice is given by the first Chern number \citep{Asboth2016short,Lohse2016pump,Nakajima2016},
the above results show that $|\nu_{1}|=1$ for the studied first band
of the temporal lattice.


\begin{thebibliography}{46}%
\makeatletter
\providecommand \@ifxundefined [1]{%
 \@ifx{#1\undefined}
}%
\providecommand \@ifnum [1]{%
 \ifnum #1\expandafter \@firstoftwo
 \else \expandafter \@secondoftwo
 \fi
}%
\providecommand \@ifx [1]{%
 \ifx #1\expandafter \@firstoftwo
 \else \expandafter \@secondoftwo
 \fi
}%
\providecommand \natexlab [1]{#1}%
\providecommand \enquote  [1]{``#1''}%
\providecommand \bibnamefont  [1]{#1}%
\providecommand \bibfnamefont [1]{#1}%
\providecommand \citenamefont [1]{#1}%
\providecommand \href@noop [0]{\@secondoftwo}%
\providecommand \href [0]{\begingroup \@sanitize@url \@href}%
\providecommand \@href[1]{\@@startlink{#1}\@@href}%
\providecommand \@@href[1]{\endgroup#1\@@endlink}%
\providecommand \@sanitize@url [0]{\catcode `\\12\catcode `\$12\catcode
  `\&12\catcode `\#12\catcode `\^12\catcode `\_12\catcode `\%12\relax}%
\providecommand \@@startlink[1]{}%
\providecommand \@@endlink[0]{}%
\providecommand \url  [0]{\begingroup\@sanitize@url \@url }%
\providecommand \@url [1]{\endgroup\@href {#1}{\urlprefix }}%
\providecommand \urlprefix  [0]{URL }%
\providecommand \Eprint [0]{\href }%
\providecommand \doibase [0]{https://doi.org/}%
\providecommand \selectlanguage [0]{\@gobble}%
\providecommand \bibinfo  [0]{\@secondoftwo}%
\providecommand \bibfield  [0]{\@secondoftwo}%
\providecommand \translation [1]{[#1]}%
\providecommand \BibitemOpen [0]{}%
\providecommand \bibitemStop [0]{}%
\providecommand \bibitemNoStop [0]{.\EOS\space}%
\providecommand \EOS [0]{\spacefactor3000\relax}%
\providecommand \BibitemShut  [1]{\csname bibitem#1\endcsname}%
\let\auto@bib@innerbib\@empty
\bibitem [{\citenamefont {{Sacha}}\ and\ \citenamefont
  {{Zakrzewski}}(2018)}]{Sacha2017rev}%
  \BibitemOpen
  \bibfield  {author} {\bibinfo {author} {\bibfnamefont {K.}~\bibnamefont
  {{Sacha}}}\ and\ \bibinfo {author} {\bibfnamefont {J.}~\bibnamefont
  {{Zakrzewski}}},\ }\href {https://doi.org/10.1088/1361-6633/aa8b38}
  {\bibfield  {journal} {\bibinfo  {journal} {Rep. Prog. Phys.}\ }\textbf
  {\bibinfo {volume} {81}},\ \bibinfo {pages} {016401} (\bibinfo {year}
  {2018})},\ \Eprint {https://arxiv.org/abs/1704.03735} {arXiv:1704.03735}
  \BibitemShut {NoStop}%
\bibitem [{\citenamefont {Guo}\ and\ \citenamefont {Liang}(2020)}]{Guo2020}%
  \BibitemOpen
  \bibfield  {author} {\bibinfo {author} {\bibfnamefont {L.}~\bibnamefont
  {Guo}}\ and\ \bibinfo {author} {\bibfnamefont {P.}~\bibnamefont {Liang}},\
  }\href {https://doi.org/10.1088/1367-2630/ab9d54} {\bibfield  {journal}
  {\bibinfo  {journal} {New J. Phys.}\ }\textbf {\bibinfo {volume} {22}},\
  \bibinfo {pages} {075003} (\bibinfo {year} {2020})},\ \Eprint
  {https://arxiv.org/abs/2005.03138} {arXiv:2005.03138} \BibitemShut {NoStop}%
\bibitem [{\citenamefont {Sacha}(2020)}]{SachaTC2020}%
  \BibitemOpen
  \bibfield  {author} {\bibinfo {author} {\bibfnamefont {K.}~\bibnamefont
  {Sacha}},\ }\href {https://doi.org/10.1007/978-3-030-52523-1} {\emph
  {\bibinfo {title} {Time Crystals}}}\ (\bibinfo  {publisher} {Springer
  International Publishing},\ \bibinfo {year} {2020})\BibitemShut {NoStop}%
\bibitem [{\citenamefont {Guo}(2021)}]{GuoBook}%
  \BibitemOpen
  \bibfield  {author} {\bibinfo {author} {\bibfnamefont {L.}~\bibnamefont
  {Guo}},\ }\href {https://doi.org/10.1088/978-0-7503-3563-8} {\emph {\bibinfo
  {title} {Phase Space Crystals}}}\ (\bibinfo  {publisher} {{IOP} Publishing},\
  \bibinfo {year} {2021})\BibitemShut {NoStop}%
\bibitem [{\citenamefont {Hannaford}\ and\ \citenamefont
  {Sacha}(2022)}]{Hannaford2022}%
  \BibitemOpen
  \bibfield  {author} {\bibinfo {author} {\bibfnamefont {P.}~\bibnamefont
  {Hannaford}}\ and\ \bibinfo {author} {\bibfnamefont {K.}~\bibnamefont
  {Sacha}},\ }\href {https://doi.org/10.1007/s43673-022-00041-8} {\bibfield
  {journal} {\bibinfo  {journal} {{AAPPS} Bulletin}\ }\textbf {\bibinfo
  {volume} {32}},\ \bibinfo {pages} {12} (\bibinfo {year} {2022})},\ \Eprint
  {https://arxiv.org/abs/2202.05544} {arXiv:2202.05544} \BibitemShut {NoStop}%
\bibitem [{\citenamefont {Guo}, \citenamefont {Marthaler},\ and\ \citenamefont
  {Sch\"on}(2013)}]{Guo2013}%
  \BibitemOpen
  \bibfield  {author} {\bibinfo {author} {\bibfnamefont {L.}~\bibnamefont
  {Guo}}, \bibinfo {author} {\bibfnamefont {M.}~\bibnamefont {Marthaler}},\
  and\ \bibinfo {author} {\bibfnamefont {G.}~\bibnamefont {Sch\"on}},\ }\href
  {https://doi.org/10.1103/PhysRevLett.111.205303} {\bibfield  {journal}
  {\bibinfo  {journal} {Phys. Rev. Lett.}\ }\textbf {\bibinfo {volume} {111}},\
  \bibinfo {pages} {205303} (\bibinfo {year} {2013})},\ \Eprint
  {https://arxiv.org/abs/1305.1800} {arXiv:1305.1800} \BibitemShut {NoStop}%
\bibitem [{\citenamefont {Sacha}(2015)}]{Sacha15a}%
  \BibitemOpen
  \bibfield  {author} {\bibinfo {author} {\bibfnamefont {K.}~\bibnamefont
  {Sacha}},\ }\href {https://doi.org/10.1038/srep10787} {\bibfield  {journal}
  {\bibinfo  {journal} {Sci. Rep.}\ }\textbf {\bibinfo {volume} {5}},\ \bibinfo
  {pages} {10787} (\bibinfo {year} {2015})},\ \Eprint
  {https://arxiv.org/abs/1502.02507} {arXiv:1502.02507} \BibitemShut {NoStop}%
\bibitem [{\citenamefont {Guo}, \citenamefont {Liu},\ and\ \citenamefont
  {Marthaler}(2016)}]{Guo2016a}%
  \BibitemOpen
  \bibfield  {author} {\bibinfo {author} {\bibfnamefont {L.}~\bibnamefont
  {Guo}}, \bibinfo {author} {\bibfnamefont {M.}~\bibnamefont {Liu}},\ and\
  \bibinfo {author} {\bibfnamefont {M.}~\bibnamefont {Marthaler}},\ }\href
  {https://doi.org/10.1103/PhysRevA.93.053616} {\bibfield  {journal} {\bibinfo
  {journal} {Phys. Rev. A}\ }\textbf {\bibinfo {volume} {93}},\ \bibinfo
  {pages} {053616} (\bibinfo {year} {2016})},\ \Eprint
  {https://arxiv.org/abs/1503.03096} {arXiv:1503.03096} \BibitemShut {NoStop}%
\bibitem [{\citenamefont {Guo}\ and\ \citenamefont
  {Marthaler}(2016)}]{Guo2016}%
  \BibitemOpen
  \bibfield  {author} {\bibinfo {author} {\bibfnamefont {L.}~\bibnamefont
  {Guo}}\ and\ \bibinfo {author} {\bibfnamefont {M.}~\bibnamefont
  {Marthaler}},\ }\href {https://doi.org/10.1088/1367-2630/18/2/023006}
  {\bibfield  {journal} {\bibinfo  {journal} {New J. Phys.}\ }\textbf {\bibinfo
  {volume} {18}},\ \bibinfo {pages} {023006} (\bibinfo {year} {2016})},\
  \Eprint {https://arxiv.org/abs/1410.3795} {arXiv:1410.3795} \BibitemShut
  {NoStop}%
\bibitem [{\citenamefont {Mierzejewski}, \citenamefont {Giergiel},\ and\
  \citenamefont {Sacha}(2017)}]{Mierzejewski2017}%
  \BibitemOpen
  \bibfield  {author} {\bibinfo {author} {\bibfnamefont {M.}~\bibnamefont
  {Mierzejewski}}, \bibinfo {author} {\bibfnamefont {K.}~\bibnamefont
  {Giergiel}},\ and\ \bibinfo {author} {\bibfnamefont {K.}~\bibnamefont
  {Sacha}},\ }\href {https://doi.org/10.1103/PhysRevB.96.140201} {\bibfield
  {journal} {\bibinfo  {journal} {Phys. Rev. B}\ }\textbf {\bibinfo {volume}
  {96}},\ \bibinfo {pages} {140201} (\bibinfo {year} {2017})},\ \Eprint
  {https://arxiv.org/abs/1706.09791} {arXiv:1706.09791} \BibitemShut {NoStop}%
\bibitem [{\citenamefont {Delande}, \citenamefont {Morales-Molina},\ and\
  \citenamefont {Sacha}(2017)}]{delande17}%
  \BibitemOpen
  \bibfield  {author} {\bibinfo {author} {\bibfnamefont {D.}~\bibnamefont
  {Delande}}, \bibinfo {author} {\bibfnamefont {L.}~\bibnamefont
  {Morales-Molina}},\ and\ \bibinfo {author} {\bibfnamefont {K.}~\bibnamefont
  {Sacha}},\ }\href {https://doi.org/10.1103/PhysRevLett.119.230404} {\bibfield
   {journal} {\bibinfo  {journal} {Phys. Rev. Lett.}\ }\textbf {\bibinfo
  {volume} {119}},\ \bibinfo {pages} {230404} (\bibinfo {year} {2017})},\
  \Eprint {https://arxiv.org/abs/1702.03591} {arXiv:1702.03591} \BibitemShut
  {NoStop}%
\bibitem [{\citenamefont {Giergiel}, \citenamefont {Miroszewski},\ and\
  \citenamefont {Sacha}(2018)}]{Giergiel2018}%
  \BibitemOpen
  \bibfield  {author} {\bibinfo {author} {\bibfnamefont {K.}~\bibnamefont
  {Giergiel}}, \bibinfo {author} {\bibfnamefont {A.}~\bibnamefont
  {Miroszewski}},\ and\ \bibinfo {author} {\bibfnamefont {K.}~\bibnamefont
  {Sacha}},\ }\href {https://doi.org/10.1103/PhysRevLett.120.140401} {\bibfield
   {journal} {\bibinfo  {journal} {Phys. Rev. Lett.}\ }\textbf {\bibinfo
  {volume} {120}},\ \bibinfo {pages} {140401} (\bibinfo {year} {2018})},\
  \Eprint {https://arxiv.org/abs/1710.10087} {arXiv:1710.10087} \BibitemShut
  {NoStop}%
\bibitem [{\citenamefont {Lustig}, \citenamefont {Sharabi},\ and\ \citenamefont
  {Segev}(2018)}]{Lustig2018}%
  \BibitemOpen
  \bibfield  {author} {\bibinfo {author} {\bibfnamefont {E.}~\bibnamefont
  {Lustig}}, \bibinfo {author} {\bibfnamefont {Y.}~\bibnamefont {Sharabi}},\
  and\ \bibinfo {author} {\bibfnamefont {M.}~\bibnamefont {Segev}},\ }\href
  {https://doi.org/10.1364/OPTICA.5.001390} {\bibfield  {journal} {\bibinfo
  {journal} {Optica}\ }\textbf {\bibinfo {volume} {5}},\ \bibinfo {pages}
  {1390} (\bibinfo {year} {2018})}\BibitemShut {NoStop}%
\bibitem [{\citenamefont {Giergiel}\ \emph {et~al.}(2019)\citenamefont
  {Giergiel}, \citenamefont {Dauphin}, \citenamefont {Lewenstein},
  \citenamefont {Zakrzewski},\ and\ \citenamefont {Sacha}}]{Giergiel2018b}%
  \BibitemOpen
  \bibfield  {author} {\bibinfo {author} {\bibfnamefont {K.}~\bibnamefont
  {Giergiel}}, \bibinfo {author} {\bibfnamefont {A.}~\bibnamefont {Dauphin}},
  \bibinfo {author} {\bibfnamefont {M.}~\bibnamefont {Lewenstein}}, \bibinfo
  {author} {\bibfnamefont {J.}~\bibnamefont {Zakrzewski}},\ and\ \bibinfo
  {author} {\bibfnamefont {K.}~\bibnamefont {Sacha}},\ }\href
  {https://doi.org/10.1088/1367-2630/ab1e5f} {\bibfield  {journal} {\bibinfo
  {journal} {New J. Phys.}\ }\textbf {\bibinfo {volume} {21}},\ \bibinfo
  {pages} {052003} (\bibinfo {year} {2019})},\ \Eprint
  {https://arxiv.org/abs/1806.10536} {arXiv:1806.10536} \BibitemShut {NoStop}%
\bibitem [{\citenamefont {Peng}\ and\ \citenamefont
  {Refael}(2018{\natexlab{a}})}]{Peng2018}%
  \BibitemOpen
  \bibfield  {author} {\bibinfo {author} {\bibfnamefont {Y.}~\bibnamefont
  {Peng}}\ and\ \bibinfo {author} {\bibfnamefont {G.}~\bibnamefont {Refael}},\
  }\href {https://doi.org/10.1103/PhysRevB.97.134303} {\bibfield  {journal}
  {\bibinfo  {journal} {Phys. Rev. B}\ }\textbf {\bibinfo {volume} {97}},\
  \bibinfo {pages} {134303} (\bibinfo {year} {2018}{\natexlab{a}})},\ \Eprint
  {https://arxiv.org/abs/1801.05811} {arXiv:1801.05811} \BibitemShut {NoStop}%
\bibitem [{\citenamefont {Peng}\ and\ \citenamefont
  {Refael}(2018{\natexlab{b}})}]{Peng2018a}%
  \BibitemOpen
  \bibfield  {author} {\bibinfo {author} {\bibfnamefont {Y.}~\bibnamefont
  {Peng}}\ and\ \bibinfo {author} {\bibfnamefont {G.}~\bibnamefont {Refael}},\
  }\href {https://doi.org/10.1103/PhysRevB.98.220509} {\bibfield  {journal}
  {\bibinfo  {journal} {Phys. Rev. B}\ }\textbf {\bibinfo {volume} {98}},\
  \bibinfo {pages} {220509} (\bibinfo {year} {2018}{\natexlab{b}})},\ \Eprint
  {https://arxiv.org/abs/1805.01896} {arXiv:1805.01896} \BibitemShut {NoStop}%
\bibitem [{\citenamefont {\v{Z}labys}\ \emph {et~al.}(2021)\citenamefont
  {\v{Z}labys}, \citenamefont {Fan}, \citenamefont {Anisimovas},\ and\
  \citenamefont {Sacha}}]{Zlabys2021}%
  \BibitemOpen
  \bibfield  {author} {\bibinfo {author} {\bibfnamefont {G.}~\bibnamefont
  {\v{Z}labys}}, \bibinfo {author} {\bibfnamefont {C.-h.}\ \bibnamefont {Fan}},
  \bibinfo {author} {\bibfnamefont {E.}~\bibnamefont {Anisimovas}},\ and\
  \bibinfo {author} {\bibfnamefont {K.}~\bibnamefont {Sacha}},\ }\href
  {https://doi.org/10.1103/PhysRevB.103.L100301} {\bibfield  {journal}
  {\bibinfo  {journal} {Phys. Rev. B}\ }\textbf {\bibinfo {volume} {103}},\
  \bibinfo {pages} {L100301} (\bibinfo {year} {2021})},\ \Eprint
  {https://arxiv.org/abs/2012.02783} {arXiv:2012.02783} \BibitemShut {NoStop}%
\bibitem [{\citenamefont {Li}\ \emph {et~al.}(2012)\citenamefont {Li},
  \citenamefont {Gong}, \citenamefont {Yin}, \citenamefont {Quan},
  \citenamefont {Yin}, \citenamefont {Zhang}, \citenamefont {Duan},\ and\
  \citenamefont {Zhang}}]{Li2012}%
  \BibitemOpen
  \bibfield  {author} {\bibinfo {author} {\bibfnamefont {T.}~\bibnamefont
  {Li}}, \bibinfo {author} {\bibfnamefont {Z.-X.}\ \bibnamefont {Gong}},
  \bibinfo {author} {\bibfnamefont {Z.-Q.}\ \bibnamefont {Yin}}, \bibinfo
  {author} {\bibfnamefont {H.~T.}\ \bibnamefont {Quan}}, \bibinfo {author}
  {\bibfnamefont {X.}~\bibnamefont {Yin}}, \bibinfo {author} {\bibfnamefont
  {P.}~\bibnamefont {Zhang}}, \bibinfo {author} {\bibfnamefont {L.-M.}\
  \bibnamefont {Duan}},\ and\ \bibinfo {author} {\bibfnamefont
  {X.}~\bibnamefont {Zhang}},\ }\href
  {https://doi.org/10.1103/PhysRevLett.109.163001} {\bibfield  {journal}
  {\bibinfo  {journal} {Phys. Rev. Lett.}\ }\textbf {\bibinfo {volume} {109}},\
  \bibinfo {pages} {163001} (\bibinfo {year} {2012})},\ \Eprint
  {https://arxiv.org/abs/1206.4772} {arXiv:1206.4772} \BibitemShut {NoStop}%
\bibitem [{\citenamefont {G\'omez-Le\'on}\ and\ \citenamefont
  {Platero}(2013)}]{Gomez-Leon2013}%
  \BibitemOpen
  \bibfield  {author} {\bibinfo {author} {\bibfnamefont {A.}~\bibnamefont
  {G\'omez-Le\'on}}\ and\ \bibinfo {author} {\bibfnamefont {G.}~\bibnamefont
  {Platero}},\ }\href {https://doi.org/10.1103/PhysRevLett.110.200403}
  {\bibfield  {journal} {\bibinfo  {journal} {Phys. Rev. Lett.}\ }\textbf
  {\bibinfo {volume} {110}},\ \bibinfo {pages} {200403} (\bibinfo {year}
  {2013})},\ \Eprint {https://arxiv.org/abs/1303.4369} {arXiv:1303.4369}
  \BibitemShut {NoStop}%
\bibitem [{\citenamefont {Messer}\ \emph {et~al.}(2018)\citenamefont {Messer},
  \citenamefont {Sandholzer}, \citenamefont {G\"org}, \citenamefont {Minguzzi},
  \citenamefont {Desbuquois},\ and\ \citenamefont {Esslinger}}]{Messer2018}%
  \BibitemOpen
  \bibfield  {author} {\bibinfo {author} {\bibfnamefont {M.}~\bibnamefont
  {Messer}}, \bibinfo {author} {\bibfnamefont {K.}~\bibnamefont {Sandholzer}},
  \bibinfo {author} {\bibfnamefont {F.}~\bibnamefont {G\"org}}, \bibinfo
  {author} {\bibfnamefont {J.}~\bibnamefont {Minguzzi}}, \bibinfo {author}
  {\bibfnamefont {R.}~\bibnamefont {Desbuquois}},\ and\ \bibinfo {author}
  {\bibfnamefont {T.}~\bibnamefont {Esslinger}},\ }\href
  {https://doi.org/10.1103/PhysRevLett.121.233603} {\bibfield  {journal}
  {\bibinfo  {journal} {Phys. Rev. Lett.}\ }\textbf {\bibinfo {volume} {121}},\
  \bibinfo {pages} {233603} (\bibinfo {year} {2018})},\ \Eprint
  {https://arxiv.org/abs/1808.00506} {arXiv:1808.00506} \BibitemShut {NoStop}%
\bibitem [{\citenamefont {Fujiwara}\ \emph {et~al.}(2019)\citenamefont
  {Fujiwara}, \citenamefont {Singh}, \citenamefont {Geiger}, \citenamefont
  {Senaratne}, \citenamefont {Rajagopal}, \citenamefont {Lipatov},\ and\
  \citenamefont {Weld}}]{Fujiwara2019}%
  \BibitemOpen
  \bibfield  {author} {\bibinfo {author} {\bibfnamefont {C.~J.}\ \bibnamefont
  {Fujiwara}}, \bibinfo {author} {\bibfnamefont {K.}~\bibnamefont {Singh}},
  \bibinfo {author} {\bibfnamefont {Z.~A.}\ \bibnamefont {Geiger}}, \bibinfo
  {author} {\bibfnamefont {R.}~\bibnamefont {Senaratne}}, \bibinfo {author}
  {\bibfnamefont {S.~V.}\ \bibnamefont {Rajagopal}}, \bibinfo {author}
  {\bibfnamefont {M.}~\bibnamefont {Lipatov}},\ and\ \bibinfo {author}
  {\bibfnamefont {D.~M.}\ \bibnamefont {Weld}},\ }\href
  {https://doi.org/10.1103/PhysRevLett.122.010402} {\bibfield  {journal}
  {\bibinfo  {journal} {Phys. Rev. Lett.}\ }\textbf {\bibinfo {volume} {122}},\
  \bibinfo {pages} {010402} (\bibinfo {year} {2019})},\ \Eprint
  {https://arxiv.org/abs/1806.07858} {arXiv:1806.07858} \BibitemShut {NoStop}%
\bibitem [{\citenamefont {Cao}\ \emph {et~al.}(2020)\citenamefont {Cao},
  \citenamefont {Sajjad}, \citenamefont {Simmons}, \citenamefont {Fujiwara},
  \citenamefont {Shimasaki},\ and\ \citenamefont {Weld}}]{Cao2020}%
  \BibitemOpen
  \bibfield  {author} {\bibinfo {author} {\bibfnamefont {A.}~\bibnamefont
  {Cao}}, \bibinfo {author} {\bibfnamefont {R.}~\bibnamefont {Sajjad}},
  \bibinfo {author} {\bibfnamefont {E.~Q.}\ \bibnamefont {Simmons}}, \bibinfo
  {author} {\bibfnamefont {C.~J.}\ \bibnamefont {Fujiwara}}, \bibinfo {author}
  {\bibfnamefont {T.}~\bibnamefont {Shimasaki}},\ and\ \bibinfo {author}
  {\bibfnamefont {D.~M.}\ \bibnamefont {Weld}},\ }\href
  {https://doi.org/10.1103/PhysRevResearch.2.032032} {\bibfield  {journal}
  {\bibinfo  {journal} {Phys. Rev. Research}\ }\textbf {\bibinfo {volume}
  {2}},\ \bibinfo {pages} {032032} (\bibinfo {year} {2020})},\ \Eprint
  {https://arxiv.org/abs/2006.01612} {arXiv:2006.01612} \BibitemShut {NoStop}%
\bibitem [{\citenamefont {Gao}\ and\ \citenamefont {Niu}(2021)}]{Gao2021}%
  \BibitemOpen
  \bibfield  {author} {\bibinfo {author} {\bibfnamefont {Q.}~\bibnamefont
  {Gao}}\ and\ \bibinfo {author} {\bibfnamefont {Q.}~\bibnamefont {Niu}},\
  }\href {https://doi.org/10.1103/physrevlett.127.036401} {\bibfield  {journal}
  {\bibinfo  {journal} {Phys. Rev. Lett.}\ }\textbf {\bibinfo {volume} {127}},\
  \bibinfo {pages} {036401} (\bibinfo {year} {2021})},\ \Eprint
  {https://arxiv.org/abs/2011.00421} {arXiv:2011.00421} \BibitemShut {NoStop}%
\bibitem [{\citenamefont {Martinez}\ \emph {et~al.}(2021)\citenamefont
  {Martinez}, \citenamefont {Giraud}, \citenamefont {Ullmo}, \citenamefont
  {Billy}, \citenamefont {Gu{\'{e}}ry-Odelin}, \citenamefont {Georgeot},\ and\
  \citenamefont {Lemari{\'{e}}}}]{Martinez2021}%
  \BibitemOpen
  \bibfield  {author} {\bibinfo {author} {\bibfnamefont {M.}~\bibnamefont
  {Martinez}}, \bibinfo {author} {\bibfnamefont {O.}~\bibnamefont {Giraud}},
  \bibinfo {author} {\bibfnamefont {D.}~\bibnamefont {Ullmo}}, \bibinfo
  {author} {\bibfnamefont {J.}~\bibnamefont {Billy}}, \bibinfo {author}
  {\bibfnamefont {D.}~\bibnamefont {Gu{\'{e}}ry-Odelin}}, \bibinfo {author}
  {\bibfnamefont {B.}~\bibnamefont {Georgeot}},\ and\ \bibinfo {author}
  {\bibfnamefont {G.}~\bibnamefont {Lemari{\'{e}}}},\ }\href
  {https://doi.org/10.1103/physrevlett.126.174102} {\bibfield  {journal}
  {\bibinfo  {journal} {Phys. Rev. Lett.}\ }\textbf {\bibinfo {volume} {126}},\
  \bibinfo {pages} {174102} (\bibinfo {year} {2021})},\ \Eprint
  {https://arxiv.org/abs/2011.02557} {arXiv:2011.02557} \BibitemShut {NoStop}%
\bibitem [{\citenamefont {Chakraborty}\ and\ \citenamefont
  {Ghosh}(2022)}]{Chakraborty2022}%
  \BibitemOpen
  \bibfield  {author} {\bibinfo {author} {\bibfnamefont {S.}~\bibnamefont
  {Chakraborty}}\ and\ \bibinfo {author} {\bibfnamefont {S.}~\bibnamefont
  {Ghosh}},\ }\href {https://doi.org/10.1016/j.dark.2022.100976} {\bibfield
  {journal} {\bibinfo  {journal} {Phys. Dark Universe}\ }\textbf {\bibinfo
  {volume} {35}},\ \bibinfo {pages} {100976} (\bibinfo {year} {2022})},\
  \Eprint {https://arxiv.org/abs/2001.04680} {arXiv:2001.04680} \BibitemShut
  {NoStop}%
\bibitem [{\citenamefont {Thouless}(1983)}]{Thouless1983}%
  \BibitemOpen
  \bibfield  {author} {\bibinfo {author} {\bibfnamefont {D.~J.}\ \bibnamefont
  {Thouless}},\ }\href {https://doi.org/10.1103/PhysRevB.27.6083} {\bibfield
  {journal} {\bibinfo  {journal} {Phys. Rev. B}\ }\textbf {\bibinfo {volume}
  {27}},\ \bibinfo {pages} {6083} (\bibinfo {year} {1983})}\BibitemShut
  {NoStop}%
\bibitem [{\citenamefont {Lohse}\ \emph {et~al.}(2016)\citenamefont {Lohse},
  \citenamefont {Schweizer}, \citenamefont {Zilberberg}, \citenamefont
  {Aidelsburger},\ and\ \citenamefont {Bloch}}]{Lohse2016pump}%
  \BibitemOpen
  \bibfield  {author} {\bibinfo {author} {\bibfnamefont {M.}~\bibnamefont
  {Lohse}}, \bibinfo {author} {\bibfnamefont {C.}~\bibnamefont {Schweizer}},
  \bibinfo {author} {\bibfnamefont {O.}~\bibnamefont {Zilberberg}}, \bibinfo
  {author} {\bibfnamefont {M.}~\bibnamefont {Aidelsburger}},\ and\ \bibinfo
  {author} {\bibfnamefont {I.}~\bibnamefont {Bloch}},\ }\href
  {https://doi.org/10.1038/nphys3584} {\bibfield  {journal} {\bibinfo
  {journal} {Nat. Phys.}\ }\textbf {\bibinfo {volume} {12}},\ \bibinfo {pages}
  {350} (\bibinfo {year} {2016})},\ \Eprint {https://arxiv.org/abs/1507.02225}
  {arXiv:1507.02225} \BibitemShut {NoStop}%
\bibitem [{\citenamefont {Nakajima}\ \emph {et~al.}(2016)\citenamefont
  {Nakajima}, \citenamefont {Tomita}, \citenamefont {Taie}, \citenamefont
  {Ichinose}, \citenamefont {Ozawa}, \citenamefont {Wang}, \citenamefont
  {Troyer},\ and\ \citenamefont {Takahashi}}]{Nakajima2016}%
  \BibitemOpen
  \bibfield  {author} {\bibinfo {author} {\bibfnamefont {S.}~\bibnamefont
  {Nakajima}}, \bibinfo {author} {\bibfnamefont {T.}~\bibnamefont {Tomita}},
  \bibinfo {author} {\bibfnamefont {S.}~\bibnamefont {Taie}}, \bibinfo {author}
  {\bibfnamefont {T.}~\bibnamefont {Ichinose}}, \bibinfo {author}
  {\bibfnamefont {H.}~\bibnamefont {Ozawa}}, \bibinfo {author} {\bibfnamefont
  {L.}~\bibnamefont {Wang}}, \bibinfo {author} {\bibfnamefont {M.}~\bibnamefont
  {Troyer}},\ and\ \bibinfo {author} {\bibfnamefont {Y.}~\bibnamefont
  {Takahashi}},\ }\href {https://doi.org/10.1038/nphys3622} {\bibfield
  {journal} {\bibinfo  {journal} {Nat. Phys.}\ }\textbf {\bibinfo {volume}
  {12}},\ \bibinfo {pages} {296} (\bibinfo {year} {2016})},\ \Eprint
  {https://arxiv.org/abs/1507.02223} {arXiv:1507.02223} \BibitemShut {NoStop}%
\bibitem [{\citenamefont {Petrides}, \citenamefont {Price},\ and\ \citenamefont
  {Zilberberg}(2018)}]{PetridesEtAl2018}%
  \BibitemOpen
  \bibfield  {author} {\bibinfo {author} {\bibfnamefont {I.}~\bibnamefont
  {Petrides}}, \bibinfo {author} {\bibfnamefont {H.~M.}\ \bibnamefont
  {Price}},\ and\ \bibinfo {author} {\bibfnamefont {O.}~\bibnamefont
  {Zilberberg}},\ }\href {https://doi.org/10.1103/PhysRevB.98.125431}
  {\bibfield  {journal} {\bibinfo  {journal} {Phys. Rev. B}\ }\textbf {\bibinfo
  {volume} {98}},\ \bibinfo {pages} {125431} (\bibinfo {year} {2018})},\
  \Eprint {https://arxiv.org/abs/1804.01871} {arXiv:1804.01871} \BibitemShut
  {NoStop}%
\bibitem [{\citenamefont {Lohse}\ \emph {et~al.}(2018)\citenamefont {Lohse},
  \citenamefont {Schweizer}, \citenamefont {Price}, \citenamefont
  {Zilberberg},\ and\ \citenamefont {Bloch}}]{LohseEtAl2018}%
  \BibitemOpen
  \bibfield  {author} {\bibinfo {author} {\bibfnamefont {M.}~\bibnamefont
  {Lohse}}, \bibinfo {author} {\bibfnamefont {C.}~\bibnamefont {Schweizer}},
  \bibinfo {author} {\bibfnamefont {H.~M.}\ \bibnamefont {Price}}, \bibinfo
  {author} {\bibfnamefont {O.}~\bibnamefont {Zilberberg}},\ and\ \bibinfo
  {author} {\bibfnamefont {I.}~\bibnamefont {Bloch}},\ }\href
  {https://doi.org/10.1038/nature25000} {\bibfield  {journal} {\bibinfo
  {journal} {Nature}\ }\textbf {\bibinfo {volume} {553}},\ \bibinfo {pages}
  {55} (\bibinfo {year} {2018})},\ \Eprint {https://arxiv.org/abs/1705.08371}
  {arXiv:1705.08371} \BibitemShut {NoStop}%
\bibitem [{\citenamefont {Zilberberg}\ \emph {et~al.}(2018)\citenamefont
  {Zilberberg}, \citenamefont {Huang}, \citenamefont {Guglielmon},
  \citenamefont {Wang}, \citenamefont {Chen}, \citenamefont {Kraus},\ and\
  \citenamefont {Rechtsman}}]{Zilberberg2018}%
  \BibitemOpen
  \bibfield  {author} {\bibinfo {author} {\bibfnamefont {O.}~\bibnamefont
  {Zilberberg}}, \bibinfo {author} {\bibfnamefont {S.}~\bibnamefont {Huang}},
  \bibinfo {author} {\bibfnamefont {J.}~\bibnamefont {Guglielmon}}, \bibinfo
  {author} {\bibfnamefont {M.}~\bibnamefont {Wang}}, \bibinfo {author}
  {\bibfnamefont {K.~P.}\ \bibnamefont {Chen}}, \bibinfo {author}
  {\bibfnamefont {Y.~E.}\ \bibnamefont {Kraus}},\ and\ \bibinfo {author}
  {\bibfnamefont {M.~C.}\ \bibnamefont {Rechtsman}},\ }\href
  {https://doi.org/10.1038/nature25011} {\bibfield  {journal} {\bibinfo
  {journal} {Nature}\ }\textbf {\bibinfo {volume} {553}},\ \bibinfo {pages}
  {59} (\bibinfo {year} {2018})}\BibitemShut {NoStop}%
\bibitem [{\citenamefont {Shirley}(1965)}]{Shirley1965}%
  \BibitemOpen
  \bibfield  {author} {\bibinfo {author} {\bibfnamefont {J.~H.}\ \bibnamefont
  {Shirley}},\ }\href {https://doi.org/10.1103/PhysRev.138.B979} {\bibfield
  {journal} {\bibinfo  {journal} {Phys. Rev.}\ }\textbf {\bibinfo {volume}
  {138}},\ \bibinfo {pages} {B979} (\bibinfo {year} {1965})}\BibitemShut
  {NoStop}%
\bibitem [{\citenamefont {Buchleitner}, \citenamefont {Delande},\ and\
  \citenamefont {Zakrzewski}(2002)}]{Buchleitner2002}%
  \BibitemOpen
  \bibfield  {author} {\bibinfo {author} {\bibfnamefont {A.}~\bibnamefont
  {Buchleitner}}, \bibinfo {author} {\bibfnamefont {D.}~\bibnamefont
  {Delande}},\ and\ \bibinfo {author} {\bibfnamefont {J.}~\bibnamefont
  {Zakrzewski}},\ }\href {https://doi.org/10.1016/S0370-1573(02)00270-3}
  {\bibfield  {journal} {\bibinfo  {journal} {Phys. Rep.}\ }\textbf {\bibinfo
  {volume} {368}},\ \bibinfo {pages} {409} (\bibinfo {year} {2002})},\ \Eprint
  {https://arxiv.org/abs/quant-ph/0210033} {arXiv:quant-ph/0210033}
  \BibitemShut {NoStop}%
\bibitem [{\citenamefont {Asb{\'o}th}, \citenamefont {Oroszl{\'a}ny},\ and\
  \citenamefont {P{\'a}lyi}(2016)}]{Asboth2016short}%
  \BibitemOpen
  \bibfield  {author} {\bibinfo {author} {\bibfnamefont {J.}~\bibnamefont
  {Asb{\'o}th}}, \bibinfo {author} {\bibfnamefont {L.}~\bibnamefont
  {Oroszl{\'a}ny}},\ and\ \bibinfo {author} {\bibfnamefont {A.}~\bibnamefont
  {P{\'a}lyi}},\ }\href {https://doi.org/10.1007/978-3-319-25607-8} {\emph
  {\bibinfo {title} {A Short Course on Topological Insulators}}},\ \bibinfo
  {series} {Lecture Notes in Physics}, Vol.\ \bibinfo {volume} {919}\ (\bibinfo
   {publisher} {Springer International Publishing},\ \bibinfo {year} {2016})\
  \Eprint {https://arxiv.org/abs/1509.02295} {arXiv:1509.02295} \BibitemShut
  {NoStop}%
\bibitem [{\citenamefont {Resta}(1999)}]{Resta1999}%
  \BibitemOpen
  \bibfield  {author} {\bibinfo {author} {\bibfnamefont {R.}~\bibnamefont
  {Resta}},\ }\href
  {https://doi.org/10.1002/(sici)1097-461x(1999)75:4/5<599::aid-qua25>3.0.co;2-8}
  {\bibfield  {journal} {\bibinfo  {journal} {Int. J. Quantum Chem.}\ }\textbf
  {\bibinfo {volume} {75}},\ \bibinfo {pages} {599} (\bibinfo {year}
  {1999})}\BibitemShut {NoStop}%
\bibitem [{\citenamefont {Soluyanov}\ and\ \citenamefont
  {Vanderbilt}(2011)}]{Soluyanov2011}%
  \BibitemOpen
  \bibfield  {author} {\bibinfo {author} {\bibfnamefont {A.~A.}\ \bibnamefont
  {Soluyanov}}\ and\ \bibinfo {author} {\bibfnamefont {D.}~\bibnamefont
  {Vanderbilt}},\ }\href {https://doi.org/10.1103/physrevb.83.035108}
  {\bibfield  {journal} {\bibinfo  {journal} {Phys. Rev. B}\ }\textbf {\bibinfo
  {volume} {83}},\ \bibinfo {pages} {035108} (\bibinfo {year} {2011})},\
  \Eprint {https://arxiv.org/abs/1009.1415} {arXiv:1009.1415} \BibitemShut
  {NoStop}%
\bibitem [{\citenamefont {Spaldin}(2012)}]{Spaldin2012}%
  \BibitemOpen
  \bibfield  {author} {\bibinfo {author} {\bibfnamefont {N.~A.}\ \bibnamefont
  {Spaldin}},\ }\href {https://doi.org/10.1016/j.jssc.2012.05.010} {\bibfield
  {journal} {\bibinfo  {journal} {J. Solid State Chem.}\ }\textbf {\bibinfo
  {volume} {195}},\ \bibinfo {pages} {2} (\bibinfo {year} {2012})},\ \Eprint
  {https://arxiv.org/abs/1202.1831} {arXiv:1202.1831} \BibitemShut {NoStop}%
\bibitem [{\citenamefont {Aligia}\ and\ \citenamefont
  {Ortiz}(1999)}]{Aligia1999}%
  \BibitemOpen
  \bibfield  {author} {\bibinfo {author} {\bibfnamefont {A.~A.}\ \bibnamefont
  {Aligia}}\ and\ \bibinfo {author} {\bibfnamefont {G.}~\bibnamefont {Ortiz}},\
  }\href {https://doi.org/10.1103/physrevlett.82.2560} {\bibfield  {journal}
  {\bibinfo  {journal} {Phys. Rev. Lett.}\ }\textbf {\bibinfo {volume} {82}},\
  \bibinfo {pages} {2560} (\bibinfo {year} {1999})},\ \Eprint
  {https://arxiv.org/abs/cond-mat/9810348} {arXiv:cond-mat/9810348}
  \BibitemShut {NoStop}%
\bibitem [{\citenamefont {Rackauckas}\ and\ \citenamefont
  {Nie}(2017)}]{JuliaDiffEq}%
  \BibitemOpen
  \bibfield  {author} {\bibinfo {author} {\bibfnamefont {C.}~\bibnamefont
  {Rackauckas}}\ and\ \bibinfo {author} {\bibfnamefont {Q.}~\bibnamefont
  {Nie}},\ }\href {https://doi.org/10.5334/jors.151} {\bibfield  {journal}
  {\bibinfo  {journal} {J. Open Res. Software}\ }\textbf {\bibinfo {volume}
  {5}},\ \bibinfo {pages} {15} (\bibinfo {year} {2017})}\BibitemShut {NoStop}%
\bibitem [{\citenamefont {Rackauckas}\ and\ \citenamefont
  {Nie}(2019)}]{JuliaDiffEq2}%
  \BibitemOpen
  \bibfield  {author} {\bibinfo {author} {\bibfnamefont {C.}~\bibnamefont
  {Rackauckas}}\ and\ \bibinfo {author} {\bibfnamefont {Q.}~\bibnamefont
  {Nie}},\ }\href {https://doi.org/10.1016/j.advengsoft.2019.03.009} {\bibfield
   {journal} {\bibinfo  {journal} {Adv. Eng. Software}\ }\textbf {\bibinfo
  {volume} {132}},\ \bibinfo {pages} {1} (\bibinfo {year} {2019})},\ \Eprint
  {https://arxiv.org/abs/1807.06430} {arXiv:1807.06430} \BibitemShut {NoStop}%
\bibitem [{\citenamefont {Kahan}\ and\ \citenamefont
  {Li}(1997)}]{JuliaDiffEqKahanLi}%
  \BibitemOpen
  \bibfield  {author} {\bibinfo {author} {\bibfnamefont {W.}~\bibnamefont
  {Kahan}}\ and\ \bibinfo {author} {\bibfnamefont {R.-C.}\ \bibnamefont {Li}},\
  }\href {https://doi.org/10.1090/s0025-5718-97-00873-9} {\bibfield  {journal}
  {\bibinfo  {journal} {Math. Comput.}\ }\textbf {\bibinfo {volume} {66}},\
  \bibinfo {pages} {1089} (\bibinfo {year} {1997})}\BibitemShut {NoStop}%
\bibitem [{\citenamefont {McLachlan}\ and\ \citenamefont
  {Atela}(1992)}]{JuliaDiffEqMcAte}%
  \BibitemOpen
  \bibfield  {author} {\bibinfo {author} {\bibfnamefont {R.~I.}\ \bibnamefont
  {McLachlan}}\ and\ \bibinfo {author} {\bibfnamefont {P.}~\bibnamefont
  {Atela}},\ }\href {https://doi.org/10.1088/0951-7715/5/2/011} {\bibfield
  {journal} {\bibinfo  {journal} {Nonlinearity}\ }\textbf {\bibinfo {volume}
  {5}},\ \bibinfo {pages} {541} (\bibinfo {year} {1992})}\BibitemShut {NoStop}%
\bibitem [{\citenamefont {Mogensen}\ and\ \citenamefont
  {Riseth}(2018)}]{JuliaOptim}%
  \BibitemOpen
  \bibfield  {author} {\bibinfo {author} {\bibfnamefont {P.~K.}\ \bibnamefont
  {Mogensen}}\ and\ \bibinfo {author} {\bibfnamefont {A.~N.}\ \bibnamefont
  {Riseth}},\ }\href {https://doi.org/10.21105/joss.00615} {\bibfield
  {journal} {\bibinfo  {journal} {J. Open Source Software}\ }\textbf {\bibinfo
  {volume} {3}},\ \bibinfo {pages} {615} (\bibinfo {year} {2018})}\BibitemShut
  {NoStop}%
\bibitem [{\citenamefont {Bezanson}\ \emph {et~al.}(2017)\citenamefont
  {Bezanson}, \citenamefont {Edelman}, \citenamefont {Karpinski},\ and\
  \citenamefont {Shah}}]{Julia}%
  \BibitemOpen
  \bibfield  {author} {\bibinfo {author} {\bibfnamefont {J.}~\bibnamefont
  {Bezanson}}, \bibinfo {author} {\bibfnamefont {A.}~\bibnamefont {Edelman}},
  \bibinfo {author} {\bibfnamefont {S.}~\bibnamefont {Karpinski}},\ and\
  \bibinfo {author} {\bibfnamefont {V.~B.}\ \bibnamefont {Shah}},\ }\href
  {https://doi.org/10.1137/141000671} {\bibfield  {journal} {\bibinfo
  {journal} {{SIAM} Rev.}\ }\textbf {\bibinfo {volume} {59}},\ \bibinfo {pages}
  {65} (\bibinfo {year} {2017})},\ \Eprint {https://arxiv.org/abs/1411.1607}
  {arXiv:1411.1607} \BibitemShut {NoStop}%
\bibitem [{\citenamefont {Lichtenberg}\ and\ \citenamefont
  {Lieberman}(1992)}]{Lichtenberg1992}%
  \BibitemOpen
  \bibfield  {author} {\bibinfo {author} {\bibfnamefont {A.}~\bibnamefont
  {Lichtenberg}}\ and\ \bibinfo {author} {\bibfnamefont {M.}~\bibnamefont
  {Lieberman}},\ }\href
  {https://link.springer.com/book/10.1007/978-1-4757-2184-3} {\emph {\bibinfo
  {title} {Regular and chaotic dynamics}}},\ Applied mathematical sciences\
  (\bibinfo  {publisher} {Springer-Verlag},\ \bibinfo {year}
  {1992})\BibitemShut {NoStop}%
\bibitem [{\citenamefont {Xiao}, \citenamefont {Chang},\ and\ \citenamefont
  {Niu}(2010)}]{Xiao2010}%
  \BibitemOpen
  \bibfield  {author} {\bibinfo {author} {\bibfnamefont {D.}~\bibnamefont
  {Xiao}}, \bibinfo {author} {\bibfnamefont {M.-C.}\ \bibnamefont {Chang}},\
  and\ \bibinfo {author} {\bibfnamefont {Q.}~\bibnamefont {Niu}},\ }\href
  {https://doi.org/10.1103/revmodphys.82.1959} {\bibfield  {journal} {\bibinfo
  {journal} {Rev. Mod. Phys.}\ }\textbf {\bibinfo {volume} {82}},\ \bibinfo
  {pages} {1959} (\bibinfo {year} {2010})},\ \Eprint
  {https://arxiv.org/abs/0907.2021} {arXiv:0907.2021} \BibitemShut {NoStop}%
\end{thebibliography}
%

\end{document}